\documentclass[aps,prb,amsmath,amssymb,twocolumn,amsfonts,longbibliography]{revtex4-2}
\usepackage{bm,xcolor,MnSymbol}
\usepackage{graphicx}
\usepackage{braket,natbib}  
\usepackage[colorlinks=true,linkcolor=magenta,urlcolor=purple,citecolor=magenta,anchorcolor=blue]{hyperref}
\usepackage[normalem]{ulem}
\usepackage{array,multirow}
\usepackage{mathtools}
\usepackage{tabularx,booktabs}
\usepackage{orcidlink}
\usepackage{esint}
\usepackage{dcolumn}
\usepackage{bm}
\usepackage{tikz-feynman}
\usepackage{appendix}
\usepackage{xcolor}
\usepackage{tikz}
\usepackage{tikzscale}
\usetikzlibrary{arrows}
\usetikzlibrary{arrows.meta}
\usetikzlibrary{decorations.pathmorphing}

\def\bk{\textbf{k}}

\def\dg{^{\dagger}}

\newcommand{\lhl}[1]{\textcolor{black}{#1}}
\newcommand{\lhlnew}[1]{\textcolor{black}{#1}}

\begin{document}

\preprint{APS/123-QED}

\title{Particle-hole Symmetric Slave Boson Method for the Mixed Valence Problem}

\author{Liam L.H. Lau\,\orcidlink{0000-0001-6603-9088}}
\email{liam.lh.lau@physics.rutgers.edu}
\affiliation{Center for Materials Theory, Department of Physics and Astronomy,
Rutgers University, 136 Frelinghuysen Rd., Piscataway, NJ 08854-8019, USA}%
\author{Piers Coleman\,\orcidlink{0000-0001-6546-5245}}%
\affiliation{Center for Materials Theory, Department of Physics and Astronomy,
Rutgers University, 136 Frelinghuysen Rd., Piscataway, NJ 08854-8019, USA}
\affiliation{Hubbard Theory Consortium, Department of Physics, Royal Holloway, University of London, Egham, Surrey TW20 0EX, UK.}

\date{\today}

\begin{abstract}
We introduce an analytic slave boson method for treating the finite $U$ Anderson impurity model. Our approach introduces two bosons to track both $Q\rightleftharpoons Q\pm1$ valence fluctuations and reduces to a single symmetric $s$-boson in the effective action, encoding all the high energy atomic physics information in the boson's kinematics, while the low energy part of the action remains unchanged across finite $U$, infinite $U$, and Kondo limits. We recover the infinite $U$ and Kondo limit actions from our approach and show that the Kondo resonance already develops in the normal state when the slave boson has yet to condense. We show that the slave rotor and $s$-boson have the same algebraic structure, and we establish a unified functional integral framework connecting the $s$-boson and slave rotor representations for the single impurity Anderson model.
\end{abstract}

\maketitle

\section{Introduction} 
The single impurity Anderson model (SIAM) and its generalizations are foundational to our understanding of strongly correlated electron systems. The model describes tightly bound $d$-electrons that experience on-site Coulomb repulsion and hybridize with dispersive conduction $c$-electrons. This model captures the physics of local moment formation in metals with rare-earth magnetic impurities in both Kondo and mixed valence regimes and also underpins Dynamical Mean Field Theory (DMFT) \cite{Georges92, DMFTRMP, DMFTRMP2} approach to strongly correlated electron systems, which maps the original lattice problem to a quantum impurity embedded in a self-consistently determined conduction electron bath. As a canonical model for strong correlation physics, it has been extended in various directions, including the periodic Anderson model for heavy fermion physics \cite{revhf1, revhf2, revhf3, revhf4, revhf5, Coleman2007} and the Anderson-Holstein model for studying the interplay between electron-electron and electron-phonon interactions \cite{sherrington_ionic_1976, riseborough_strong_1987, hewson1979, hewsonnewns1980, hewson_scaling_1981, hewson_numerical_2001, hewson_numerical_2010,monreal_equation_2009, hotta_enhanced_2007, laakso_functional_2014, eidelstein, lauoscillate2025}.

For an impurity with $Q$ $d$-electrons (where $Q$ is less than the total number of $d$-electron flavors $N$), the SIAM describes the low energy formation of a free local moment that becomes spin-screened by the conduction sea, creating a spin singlet bound state- known as the Kondo effect. This bound state formation represents the strong coupling fixed point of the Kondo and the infinite $U$ Anderson models, where valence fluctuations $Q\rightleftharpoons Q \pm 1$, corresponding to lower and upper Hubbard bands, have been integrated out of the theory. However, depending on the degree of mixed valence, these valence fluctuations may influence the low energy physics.

The ideas from the SIAM has also recently been applied to two dimensional van der Waals materials such as magic-angle twisted bilayer graphene where the Bistritzer-MacDonald model was mapped to a topological heavy fermion model \cite{song21, shi_heavy_2022, Lau23} and the key energy scales are set by the impurity limit, allowing methods and ideas from the heavy fermion community to be applied to MATBG \cite{lau2025, sankarkondo, tsvelikkondo, tsveliksymmetric, matbgdmft, dzeromixedvalence, herzogarbeitman2025heavyfermionsefficientrepresentation}. Another example is gate-tunable $AB$-stacked $\textrm{MoTe}_2/\textrm{WSe}_2$ moir\'e bilayers \cite{Mak23} which exhibits physics spanning from Kondo to mixed valence regimes \cite{guerci23}. 
\begin{figure*}[t!]
    \centering
    \includegraphics[width = 0.8\linewidth]{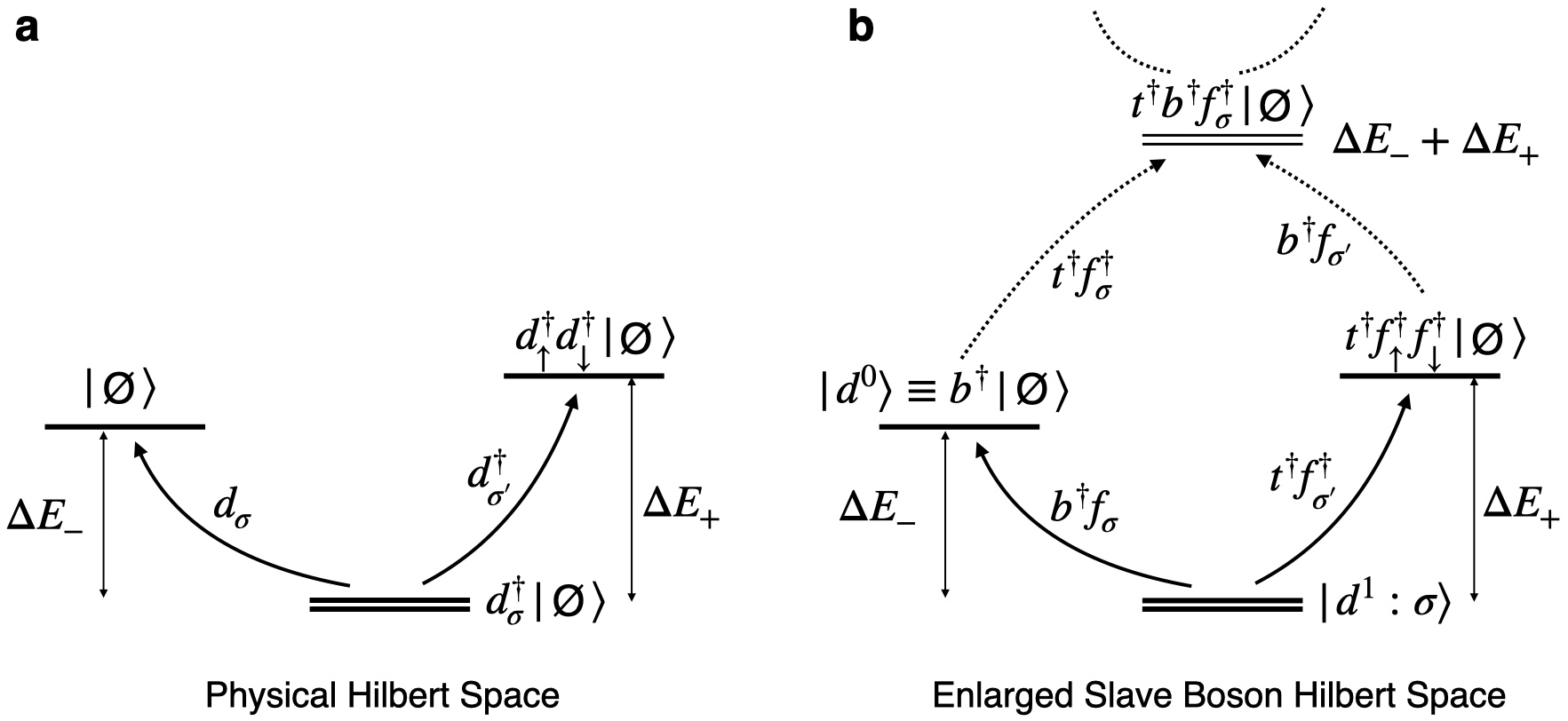}
    \caption{Schematic of the atomic energy levels of the single impurity Anderson model shown in (a) the physical $d$-electron Hilbert space and (b) the enlarged slave boson Hilbert space. In the physical Hilbert space, $\Delta E_-$ and $\Delta E_+$ represent the excitation energies to remove or add a $d$-electron to the singly occupied state $d\dg_\sigma \vert \emptyset \rangle$, respectively. The slave boson representation accurately captures the physical atomic energy levels when the excitation energies $\Delta E_{\pm}$ are sufficiently large to suppress transitions to fictitious states (dashed lines).}
    \label{fig:fig1}
\end{figure*}
Capturing this broad spectrum of low energy behavior presents a challenge for theoretical frameworks (``impurity solvers''). When the $Q$ is a fixed integer, 
the upper and lower Hubbard bands corresponding to the impurity valence fluctuations $Q\rightleftharpoons Q\pm1$ can be integrated out of the theory giving rise to the Kondo model. The Kondo limit can be exactly solved using Bethe ansatz \cite{AndreiRMP} and treated approximately using mean-field theory. Mean-field theory \lhl{captures} the Kondo resonance physics of the strong-coupling limit, describing an Abrikosov-Suhl resonance with a width of order the Kondo temperature $T_K$, corresponding to the inverse lifetime of the bound state.
To go beyond the static mean-field theory to describe both the Kondo and mixed valence dynamics, several approaches have been developed which generalize the slave boson approach. Here we recapitulate the two most prominent: slave bosons and slave rotors. 

The original slave boson representation \cite{slaveb, read83_v2,coleman1987x} describes the projected
Hubbard operators \cite{hubbardops}, \lhl{$X_{\sigma 0} \equiv d\dg_{\sigma}P_0$} relevant to infinite-U Anderson or  Hubbard models as  
\begin{equation}\label{eq:originalslaveboson}
    d\dg_\sigma P_0 \rightarrow f\dg_\sigma b,
\end{equation}
where $P_0= \vert d^0\rangle \langle d^0 \vert $, which projects into the empty state, is replaced by a boson $b$.
This factorization  into canonical fields converts the non-holonomic constraint $n_d\leq 1$ of the infinite $U$ Hubbard model into a holonomic constraint accessible to field-theory such that  
\begin{equation}\label{eq:slavebosonconstraint}
    Q = n_b + n_f, \qquad [H,Q]=0
\end{equation}
where $n_b = b\dg b$ and $n_f = \sum_\sigma f\dg_\sigma f_\sigma$ are the number of bosons and spinons respectively. 
Since $Q$ commutes with the Hamiltonian, it generates a gauge symmetry that
polices the constraints.  By choosing $Q=1$, we recover the infinite $U$ models. In $SU(N)$ models, in which the spin component $\sigma \in \{1, 2 \dots N\}$ can have $N$ possible values, $Q$ can have any integer value between $1$ and $N-1$. Increasing $N$, keeping $Q/N=q$ fixed generates a family of models \lhl{with} a controlled large $N$ expansion. 
The slave boson method preserves the commutation algebra of Hubbard operators, while enabling  analytic calculations of the fluctuations around mean-field theory. However its great draw-back is that  it treats valence fluctuations asymmetrically, only describing charge fluctuations to the lower valence state $d^1 \rightleftharpoons d^0$  relevant to infinite $U$ models. 

\lhl{Subsequent work by Kotliar and Ruckenstein \cite{kotliarruckenstein} generalized the slave boson method to include double occupation by introducing four bosons to keep track of the two singly occupied states, holon, and doublon. Although able to reproduce the Gutzwiller projection, analytic treatments of the fluctuations are cumbersome. By contrast, slave rotors \cite{physrevb.66.165111, florens_prb2004}, employ the phase $\theta$, dual to the impurity charge as the collective variable.}

The aim of this paper is to build on the boson and rotor approaches, introducing a bosonic representation that combines an economy of calculation with a symmetric treatment of valence fluctuations. The key representation  
\begin{equation}\label{eq:twobosonrep}
    d\dg_\sigma \rightarrow f\dg_\sigma \left(b + t\dg\right),
\end{equation}
involves a spinon $f\dg_{\sigma}$ and two bosonic partons, ``top'' ($t$) and ``bottom''($b$), \lhl{that} keep track of addition and removal of charge, respectively. This representation preserves the conserved
charge $Q = n_f +b\dg b - t\dg t$ which is set to an integer value in the physical subspace \cite{coleman1987x}. We note that this representation of the atomic problem expands the Hilbert space (see Fig. \ref{fig:fig1}) and is therefore not entirely faithful. However, when the atomic excitations $\Delta E_\pm$ are sufficiently large, fictitious states associated with multiple top or bottom bosons are suppressed, and the Feynman diagrams for single top or bottom boson exchange accurately track the valence fluctuations. 

 

We show that the antisymmetric combination $b-t\dg$ decouples from the theory, leading to an effective theory for the symmetric combination $s = b + t \dg$, allowing a reduction of the theory  to a \textit{single} symmetric $s$-boson theory.  At the end of this procedure, the 
original slave boson approach can be transformed into the $s$-parton formulation by replacing $b\rightarrow s$ in the hybridization and 
\begin{equation}
    b\dg (\partial_\tau + \lambda ) b \rightarrow  \frac{1}{U}\ s\dg \left[ \left(\frac{U}{2}\right )^2 - {(\partial _\tau + \lambda )^2}\right ] s.
\end{equation}
This term replaces the single pole at $\omega = \lambda$ by a double pole at  $\omega = \lambda \pm U/2$ in the $s$-boson.  With this economic transformation the slave boson approach can be upgraded to include symmetric valence fluctuations without having to recalculate the fermionic RPA bubbles \lhl{in a controlled large-$N$ expansion}. We further show that the infinite $U$ and Kondo limits can be recovered from our approach. 

This article is organized as follows. In section \ref{sec:method} we introduce a representation of the impurity fermionic operators in terms of a spinon, top $t$ and bottom $b$ boson, to keep track of positive and negative valence fluctuations. We show that in the finite $U$ multiorbital SIAM, the representation can be reduced to a single $s$-boson which symmetrically captures the lower and upper Hubbard band physics. In section \ref{sec:applicationtoSIAM}, we present the result of our single $s$-boson approach to the SIAM, \lhl{showcasing calculations of the} the local spectral function and charge susceptibility. In the discussion section \ref{sec:discussion}, we \lhl{explore} the algebraic relationship between our $s$-boson formulation and slave rotor approach, demonstrating the equivalence of their actions for the SIAM under a unified functional integral approach when we forgo the $s$-boson's large-$N$ expansion by normalizing the $s$-boson algebra. We end with a short discussion on the treatment of strong mixed valence where fluctuations into the fictitious states with multiple $t$ and $b$ bosons (Fig. \ref{fig:fig1}) are not energetically suppressed and suggest future extensions of our method.

\section{
Method}\label{sec:method}

\subsection{Atomic and Single Impurity Anderson Model}
We first recapitulate the physics of the SIAM, starting with the atomic limit for an $N$-fold degenerate atomic level with a $SU(N)$ symmetric onsite Coulomb interaction,
\begin{equation}\label{eq:atomicmodel}
H_{A} = \epsilon_d n_d +  \frac{U}{2} \left( n_d - N/2\right)^2.
\end{equation}
{Here, $\epsilon_d$  is $d$-level position at half filling, and the Coulomb interaction is defined relative to half-filling, $n_d = \sum_\sigma d\dg_\sigma d_\sigma$,  where $\sigma =  1, 2, \dots N$ labels the spin and orbital degrees of freedom.
}

In the SIAM, electrons can tunnel into and out \lhl{of} the atomic level \eqref{eq:atomicmodel} to a conduction sea with dispersion $\epsilon_{\bk}$,
\begin{equation}\label{eq:bath}
   H_{c} =  \sum_{\bk \sigma} \epsilon_{\bk} c\dg_{\bk \sigma} c_{\bk \sigma}, 
\end{equation}
through the hybridization
\begin{equation}\label{eq:hybridization}
    H_{\text{hyb}} = \frac{V_0}{\sqrt{N}} \sum_{\bk \sigma} \left(c\dg_{\bk \sigma} d_{\sigma} + \text{h.c.} \right).
\end{equation}
In general, the hybridization can be momentum dependent, but in this work we will take the simplified scenario where the hybridization is momentum independent and the conduction bath is featureless and has a constant density of states $\rho_c$. The full SIAM Hamiltonian is,
\begin{equation}\label{eq:SIAM}
   H =  H_{\text{c}} + H_{A}  + H_{\text{hyb}} .
\end{equation}
The hybrdization between the atom and a conduction bath allows for the valence of the atom to fluctuate ($Q \rightleftharpoons Q \pm1$) around each integer $Q$ with ionization energies,
\begin{eqnarray}\label{eq:ionizationenergies}
    \Delta E^Q_{+}  &=& E_{Q+1} - E_Q = \tilde{\epsilon}_d + U/2, \cr 
    \Delta E^Q_{-} &=& E_{Q-1} - E_Q = -\tilde{\epsilon}_d + U/2.
\end{eqnarray}
where we have introduced the notation
\begin{eqnarray}
    \tilde{\epsilon}_d = \epsilon_d + U (Q-N/2)
\end{eqnarray} for the $d$-level position at doping away from half filling. The atomic ground-state has $n_d=Q$, provided $|\tilde\epsilon_d|> U/2$, i.e provided $\epsilon_d$ lies within $U/2$ of $U(Q-N/2) $.   
 The upper and lower Hubbard band positions \lhl{correspond to} resonances in the d-electron spectral function at energies
 \begin{equation}\label{eq:ulhbs}
     \omega_+ = \Delta E^Q_{+}, \quad\omega_- = -\Delta E^Q_{-} ,
 \end{equation}
 which are separated by the onsite Coulomb scale $U$.
\subsection{Slave Boson Representation of SIAM} 
 Unlike the infinite $U$ Anderson model, we now need to keep track of charge fluctuations due to \textit{adding} and \textit{removing} an electron from the atom. We do so by first rewriting the physical $d$-electron as the product of a spinon $f_\sigma$ which lives in a constrained Hilbert space and introduce two bosons, $t$ and $b$, the ``top'' and ``bottom'' bosons which keep track of the charge fluctuations into the $Q+1$ and $Q-1$ valence states, respectively.  We then write
 \begin{eqnarray}
    d\dg_{\sigma} &\rightarrow& f\dg_{\sigma} (b + t\dg).
 \end{eqnarray}

The inclusion of these slave bosons expand the Hilbert space, therefore a softened constraint is required to fix it back to be the physical Hilbert space,
\begin{equation}\label{eq:constraint}
   n_{f} + n_b - n_t = Q. 
\end{equation}
$Q$ describes a fixed gauge charge of the mixed valent system, in which the top and bottom bosons have 
opposite  gauge charge.  
Within the physical subspace constrained by Eq. \ref{eq:constraint}, the Hamiltonian,
\begin{eqnarray}\label{eq:enlargedhilberthammy}
   \tilde{H} &=& H_{c} + \frac{V_0}{\sqrt{N}} \sum_{\bk \sigma} \left(c\dg_{\bk \sigma} f_{\sigma} \left(b\dg + t \right) + \text{h.c.} \right) \cr &+& \Delta E^{Q}_-  \;b\dg b + \Delta E^{Q}_- \; t\dg t .
\end{eqnarray}
provides a low-energy description of the single-impurity Anderson model.
As an important initial check, we note that if the conduction bath is particle-hole symmetric, i.e  $\epsilon_\bk = -\epsilon_{-\bk}$, then under the particle-hole transformation
\begin{eqnarray}
c_k&\rightarrow &c\dg_{-k},\qquad f\rightarrow f\dg, \qquad t \rightarrow b\dg \cr
\lambda &\rightarrow &- \lambda, \qquad Q\rightarrow N-Q, \qquad \epsilon_d \rightarrow - \epsilon_d
\end{eqnarray}
the action $H + \lambda(n_f+n_b-n_t-Q)$ is invariant. 

The  partition function in the Gibb's ensemble of definite $Q$ is then written 
\begin{equation}
Z_Q = \int _{\lambda_0}^{\lambda_0+ 2 \pi i k_B T} \frac{d\lambda}{2 \pi i k_B T} {\rm Tr}
\left[e^{- \beta \bigl(H + \lambda (n_f+n_b-n_t- Q)\bigr)}\right]
\end{equation}
where the chemical potential $\lambda$, integrated along the imaginary axis from $\lambda = \lambda_0$ to $\lambda = \lambda_0 + 2 \pi i k_B T$ imposes the constraint on the gauge charge\eqref{eq:constraint}. By extending the $\lambda$ integral along the imaginary axis, we can then rewrite the partition function as a path integral,  
\begin{eqnarray} 
Z_Q &= &\int D[ B,F] \exp{\left[-\int _0^\beta d\tau L\right]},\cr
L &=& L_F +L_{\text{hyb}} + L_B,
\end{eqnarray}
where $D[B]\equiv D[\bar b, b, \bar t, t, \lambda ]$ describes the measure of integration over the slave bosons  $b$, $t$ and the constraint field $\lambda$, while $D[F] = D[c\dg,c,f\dg, f]$ denote the measures of integration over the conduction and f-electrons. \begin{equation}\label{eq:SAVF}
    L_F = \sum _{\bk, \sigma} c\dg_{\bk \sigma }(\partial_{\tau}+ \epsilon_{\bk})c_{\bk \sigma}+\sum_\sigma f\dg_{\sigma} (\partial_{\tau} + \lambda) f_{\sigma}
\end{equation}
is the Lagrangian for the fermions,  
\begin{eqnarray}\label{eq:SAV0}
    L_{\text{hyb}}= \frac{V_0}{\sqrt{N}}\sum_{\bk \sigma} \left(c\dg_{\bk \sigma} f_{\sigma} \left(\bar{b} + t \right) + \text{H.c.} \right), 
\end{eqnarray}
describes the hybridization and
\begin{eqnarray}\label{eq:SA00}
L_B&=&  \bar{b} (\partial_{\tau} + \Delta E^Q_- + \lambda) b +  \bar{t} (\partial_{\tau} + \Delta E^Q_+ - \lambda) t \cr &-& \lambda Q
\end{eqnarray}
is the Lagrangian for  the bosons.  The constraint field $\lambda$ couples to the f-electron in \eqref{eq:SAVF}, and to the bottom and top bosons $b$ and $t$ in \eqref{eq:SA00} with opposite charge.
\\

\vskip 0.2truein
\subsection{Effective Action}

By combining the bottom  with the conjugated top boson in symmetric and antisymmetric combinations
\begin{equation}\label{eq:newbosons}
   s = (b + \bar t), \quad a = (b - \bar t),
\end{equation}
we can  rewrite the hybridization and atomic Lagrangians \eqref{eq:SAV0} and \eqref{eq:SA00} as,
\begin{widetext}
    \begin{eqnarray}\label{eq:SA1}
    L_{\text{hyb}}&=& \frac{V_0}{\sqrt{N}}\sum_{\bk \sigma} \left(c\dg_{\bk \sigma} f_{\sigma} {\color{black} \bar{s}}  + \text{H.c.} \right), \cr
    L_B &=& \left[\frac{1}{4}(\bar{s} + \bar{a}) (\partial_{\tau} + \frac{U}{2} - \tilde{\epsilon}_d+ \lambda )(s + a) +   {\frac{1}{4}}(s - a )(\partial_{\tau}  + \frac{U}{2} + \tilde{\epsilon}_d -\lambda) (\bar{s} - \bar{a}) - \lambda Q  \right],
\end{eqnarray}
The antisymmetric boson $a$ decouples from the fermions, and we can integrate it out 
(see Appendix \ref{sec:appendixeffectiveaction} for details), so that the bosonic action 
\begin{eqnarray}\label{eq:SA2}
   L_B  =  \frac{1}{U}\bar{s}
   \biggl(\frac{U^2}{4\ }- (\partial_{\tau}  + \lambda -\tilde{\epsilon}_d )^2 \biggr) s - \lambda Q .
\end{eqnarray}
now contains 
two poles at $\omega = \lambda-\tilde{\epsilon}_d\pm U/2$,
corresponding to the upper and lower Mott bands.  The complete action for our $s$-boson reformulation is then
\begin{equation}\label{eq:Csummary}
  S= \int_0^{\beta}d\tau \left[
     \frac{1}{U}\bar{s}
   \biggl(\frac{U^2}{4\ }- (\partial_{\tau}  + \lambda -\tilde{\epsilon}_d )^2 \biggr) s + \sum _{\bk, \sigma} c\dg_{\bk \sigma }(\partial_{\tau}+ \epsilon_{\bk})c_{\bk \sigma}+\sum_\sigma f\dg_{\sigma} (\partial_{\tau} + \lambda) f_{\sigma} +\frac{V_0}{\sqrt{N}}\sum_{\bk \sigma} \left(c\dg_{\bk \sigma} f_{\sigma} {\color{black}
   \bar s}  + \text{H.c.} \right) - \lambda Q\right].
\end{equation}
The economy of our approach is thus revealed, for apart from the modified bosonic action, i.e the modified atomic physics incorporating the upper and lower Mott bands, this is identical to the original slave boson formulation. This action has a controlled large-$N$ expansion due to $\vert s \vert \sim \mathcal{O}(\sqrt{N})$, thus every term in \eqref{eq:Csummary} scales with $N$.

\lhl{
\subsection{Atomic Limit}
Before demonstrating that our method recovers the infinite $U$ and Kondo limits of the Anderson model and performing mean-field theory with Gaussian fluctuations for the finite $U$ case, we first show that our method exactly captures the atomic limit at $T=0$. In this atomic limit ($V_0 = 0$, ignoring the conduction bath), the action \eqref{eq:Csummary} becomes,
\begin{equation}\label{eq:atomicS}
  S_{\text{atom}}= \int_0^{\beta}d\tau \left[
     \frac{1}{U}\bar{s}
   \biggl(\frac{U^2}{4\ }- (\partial_{\tau}  + \lambda -\tilde{\epsilon}_d )^2 \biggr) s +\sum_\sigma f\dg_{\sigma} (\partial_{\tau} + \lambda) f_{\sigma} - \lambda Q\right].
\end{equation}
Let us now consider the average impurity occupancy $\langle \hat Q \rangle = \langle \sum_\sigma d\dg_\sigma d_\sigma \rangle$ on the position of the atomic level $\epsilon_d$.
The average constraint equation is obtained by varying the atomic action with respect to $\lambda$,
\begin{eqnarray}\label{eq:minimizewrtlambdaatomic}
   \langle n_{f} \rangle   =Q - 2\frac{ \left(\tilde{\epsilon}_d-\lambda\right) }{U} \langle \bar s s \rangle.
\end{eqnarray}
In the limit $T \rightarrow 0$, we find that the constraint vanishes linearly with temperature $\lambda = T\ln{\left(N/Q - 1 \right)}$. The expectation $\langle \bar s s \rangle = 0$ at $T = 0$ holds only if $\vert \tilde \epsilon_d \vert < U/2$, then we can immediately write down the average impurity charge $\langle \hat{Q} \rangle$ as a function of $\epsilon_d$,
\begin{equation}
    \langle \hat{Q} \rangle = \langle n_d \rangle = \langle n_f \rangle = Q, \quad -\frac{U}{2} \left(2 Q + 1 - N\right) < \epsilon_d < -\frac{U}{2}\left(2Q -1 - N\right),
\end{equation}
corresponding to the Coulomb staircase provided that the expansion around the correct $Q$ is made.}
\lhl{
We further show that the spectral function of the physical electron in the atomic \lhlnew{limit} is correct. The Green's function for the physical electron is $G_d(\tau) \equiv - \langle d_\sigma (\tau) d_\sigma\dg(0) \rangle$, such that the exact physical electron spectral function in the atomic limit at $T=0$ is given by Appendix \ref{sec:atomiclimitapp},
\begin{eqnarray}
    A^{at}_d(\omega) &=& \frac{1}{\pi}\left[n_B(\lambda  - \omega ) + \frac{Q}{N} \right]\,  D^{''}_s(\lambda - \omega) \cr &=& \left(1 - \frac{Q}{N} \right)\, \delta\left(\omega - \Delta E^{Q}_+ \right) + \frac{Q}{N} \,\delta\left(\omega + \Delta E^Q_-\right),
\end{eqnarray}
where \lhlnew{$\Delta E^{Q}_\pm$ is defined in Eq. \eqref{eq:ionizationenergies}}, which are the precursors of the Mott bands in a solid. 
} 

\subsection{Recovery of the infinite $U$ and Kondo limits}
Let us now verify that the effective action \eqref{eq:Csummary} recovers the well known infinite $U$ and  Kondo limits of the Anderson model. We first factorize 
\begin{eqnarray}\label{eq:SA2X}
   L_B & = &
   \frac{1}{U}\bar{s}
   \left[\left(\frac{U}{2}+(\partial_\tau + \lambda - \tilde{\epsilon}_d)\right)\left(\frac{U}{2}-(\partial_\tau + \lambda - \tilde{\epsilon}_d)\right) \right] s- \lambda Q \cr
    & = &
   \frac{1}{U}\bar{s}
   \biggl[\left(\partial_\tau + \lambda - E_d\right)\left( U+E_d-\partial_\tau - \lambda \right) \biggr] s- \lambda Q ,
\end{eqnarray}
where we have redefined $E_d= \tilde{\epsilon}_d-U/2$. The infinite $U$ limit is obtained by taking $U\rightarrow\infty$,
maintaining $E_d$ fixed, allowing the second term of \eqref{eq:SA2X} to be replaced by $U$, recovering the action
of the infinite $U$ Anderson model, 
\begin{equation}\label{eq:SA2Z}
L_B^{U=\infty} = \bar s\left(\partial_\tau + \lambda - E_d\right)s- \lambda Q .
\end{equation}

The Kondo limit occurs when the positions of the upper and lower Mott bands exceed the band-width.  In this situation, the high-frequency charge dynamics of the $s$-boson are eliminated, and since the $d$-level position is far smaller than the Mott band energies, we can set the frequency  $\partial_\tau +\lambda \rightarrow 0$ in $L_B$. Rescaling $s$ as a hybridization field by writing $\frac{V_0}{\sqrt{N}} s = V$, the resulting Kondo action is,
\begin{eqnarray}\label{eq:SA2Kondo}
   S_K  &=& \int_0^{\beta} d \tau \left[\sum _{\bk, \sigma} c\dg_{\bk \sigma }(\partial_{\tau}+ \epsilon_{\bk})c_{\bk \sigma}+\sum_\sigma f\dg_{\sigma} (\partial_{\tau} + \lambda  ) f_{\sigma}   +  \sum_{\bk \sigma} \left(\bar{V}c\dg_{\bk \sigma} f_{\sigma}  + \text{H.c.} \right)+ N \frac{\bar V V}{J} - \lambda Q \right],
\end{eqnarray}

This recovers the well-known
Newns-Read formulation \cite{Read1983, read_new_1983} of the Coqblin-Schrieffer model \cite{Coqblin69}, but now, with the well-known Schrieffer-Wolff expression for the 
Kondo coupling constant, 
\begin{equation}\label{sw}
    J = \frac{V_0^2}{U/2 - \tilde \epsilon _d }+ \frac{V_0^2}{U/2 + \tilde \epsilon _d }
\end{equation}

\end{widetext}

\subsection{Mean-field Theory}

By integrating \eqref{eq:SA2} over the fermionic fields, we obtain the effective action in terms of $\lambda$ and $s$,
\begin{eqnarray}\label{eq:effaction}
   Z = \int D[\lambda, s] \exp{\left(-S_E[\lambda, s]\right)}. 
\end{eqnarray}
Seeking a static mean-field theory, varying $S_E$ with respect to $s$ we obtain
\begin{eqnarray}\label{eq:minimizewrts}
     \frac{1}{U}\left(\frac{U^2}{4}-(\lambda - \tilde \epsilon_d)^2\right)s + \frac{V_0}{\sqrt{N}}\sum_{\bk \sigma} \langle c\dg_{\bk \sigma} f_\sigma\rangle_{s } = 0,
\end{eqnarray}
while varying $S_E$ with respect to $\lambda$, the mean-field constraint \eqref{eq:constraint} now becomes,
\begin{eqnarray}\label{eq:minimizewrtlambda}
   \langle n_{f} \rangle +2\frac{ \left(\tilde{\epsilon}_d-\lambda\right) }{U}|s|^2  =Q.
\end{eqnarray}
The second term in this expression is identified as the expectation value of $|b|^2 - |t|^2$, denoting the deviation from integral valence. 

We now present some analytical results for a single impurity Anderson model with a conduction sea with a constant density of states $\rho_c$. If we assume a constant density of conduction states $\rho_c$ with a band-width $2D$, then the mean-field impurity free energy is given by (see Appendix \ref{sec:meanfieldanalytics} for details),
\begin{equation}\label{eq:analyticmfF}
    F = -TN \sum_{\omega_n < D_c} \ln{\left[\lambda + i \tilde \Delta_n - i\omega_n\right]} + \frac{\vert s \vert^2 V_0^2}{J(\lambda)} - \lambda Q .
\end{equation}
Here $D_c = \text{min}\left(D, U/2\right)$, the high-energy cutoff, depends on the relative size of the bandwidth $D$ and the Coulomb interaction $U$, while $i \tilde\Delta_n = i \pi \rho_c V_0^2 \vert \tilde s\vert^2 \text{sgn}(n)$ is the large-cutoff approximation for the f-electron self energy, where $\tilde s = s/\sqrt{N}$. Lastly, the  $\lambda$-dependent Kondo coupling has the Schrieffer-Wolff form,
\begin{equation}\label{eq:KondoJgeneralized}
    J(\lambda) = \frac{V_0^2}{U/2 + \lambda - \tilde \epsilon _d }+ \frac{V_0^2}{U/2 - \lambda + \tilde \epsilon _d }
\end{equation}
Within mean-field theory, the Kondo effect has two effects, firstly to float the l-level $\lambda$ at the Fermi level, and secondly to renormalize the resonant level hybridization width from the large-$N$ bare hybridization width $\Delta = \pi \rho_c V_0^2$ to 
$\tilde \Delta = \pi \rho_c V_0^2 \vert \tilde s\vert^2$.

Using the complex $d$-level position  
\begin{eqnarray}\label{cfpos}
    \xi = \lambda + i \tilde \Delta,
\end{eqnarray}whose real and imaginary parts respectively represent the position and width of the resonant level, the mean-field Free energy can be written as 
\begin{equation}\label{MFFE}
    F= N {\rm Im}\left[2 i T \ln \tilde \Gamma(\xi)- \frac{\xi}{\pi}\ln \frac{D_c}{2\pi i T}\right]+  \frac{\vert s \vert^2 V_0^2}{J(\lambda)}- \lambda Q
\end{equation}
where  
$
    \tilde{\Gamma}(\xi) = \Gamma\left(\frac{1}{2} + \frac{\xi}{2 \pi i T} \right),
$
is a short-hand for the gamma function.  We can also rewrite the free energy  in the compact form $F= N {\rm Im} F_c$, where
\begin{eqnarray}\label{MFFEX}
    F_c= 2 i T \ln \tilde \Gamma(\xi)- \frac{\xi}{\pi}\ln \frac{T_K(\lambda)e^{ i \pi q}}{2\pi i T},
\end{eqnarray}
where 
\begin{eqnarray}\label{eq:TKlargeN}
 &&T_K(\lambda) = D_c \exp{\left[-\frac{1}{ \rho_c J(\lambda)}\right]} \cr &=& D_c \exp{\left[-\frac{\pi}{ \Delta} \left(\frac{1}{U/2 + \lambda - \tilde \epsilon_d} + \frac{1}{U/2 - \lambda + \tilde \epsilon_d} \right)^{-1}\right]}
\end{eqnarray}
is the $\lambda$-dependent Kondo temperature and $q=\frac{Q}{N}$. 

In the Kondo limit, we can neglect the $\lambda$ dependence of the Kondo coupling constant, setting $J = J(\lambda=0)$. In this limit, we obtain the saddle point equations by differentiating \eqref{MFFEX} with respect to $\xi$. Setting $\partial F_c/\delta \xi = 0$,   we obtain 
\begin{equation}
    \tilde \psi(\xi) = \ln \left(\frac{T_K e^{i \pi q}}{2 \pi  i T}\right),
\end{equation}
where  we  denote,
$
    \tilde{\psi}(\xi) = \psi\left(\frac{1}{2} + \frac{\xi}{2 \pi i T} \right),
$
and $\psi(z) = d \ln \Gamma(z)/dz $ is the digamma function. $T_K = T_K(\lambda)\vert_{\lambda =0} =D_c \exp{\left[-\frac{1}{ \rho_c J}\right]} $ is the Kondo temperature.  At low temperature, we can approximate $\tilde \psi (z) = \ln (\frac{z}{2 \pi i T})$, so we obtain simply $\xi = \lambda + i \tilde \Delta = T_K e^{i \pi q}$, recovering the large $N$ limit of the SU(N) Kondo model, but with the finite $U$  Schrieffer-Wolff form of the coupling constant $J$\eqref{sw}.  

In the strongly  mixed valent regime, we must take account of the $\lambda$ dependence of the Kondo coupling constant.  
In this case, the mean field  equation from varying  \eqref{MFFE} with respect to $\tilde s$ can be written as,
\begin{equation}
    \,  \text{Re}\left[\tilde \psi (\xi) - \ln\left(\frac{  T_K(\lambda)}{2\pi T}\right) \right] =0.
\end{equation}
  The saddle-point condition $\partial F/\partial \lambda=0 $ is
\begin{equation}\label{eq:flatdosminimizewrtlambda}
    \frac{1}{\pi}\,  \text{Im}\left[\tilde \psi (\xi) + \frac{i \pi}{2} \right] + 2\frac{ \left(\tilde{\epsilon}_d-\lambda\right) }{U}|\tilde s|^2=q,
\end{equation}
where the second term is recognized as the deviation from integral valence found in \eqref{eq:minimizewrtlambda}.
We can combine the two equations in complex form,
\begin{equation}\label{comp}
    \tilde\psi(\xi) - \ln{\left(\frac{T_K(\lambda) e^{i \pi q}}{2 \pi i T}\right)} = - 2 \pi i \frac{\tilde \epsilon_d-\lambda }{U} \vert \tilde s \vert^2.
\end{equation}
The transition temperature $T_c$ into the mean-field Kondo phase as a function of $d$-level can be found by solving \eqref{comp} in the limit where the $s$-boson has not yet condensed, $\tilde s = 0$, and the constraint $\lambda$ is thermally determined by the Fermi function, $f(\lambda) = q$, where $f(\lambda)= 1/ (e^{\beta \lambda} +1) $ is the Fermi function and $q = Q/N$ is the impurity filling factor. This can be rewritten as,
\begin{equation}\label{eq:Tcsolve}
    \text{Re}\left[\psi \left(\frac{1}{2} + \frac{\log{\left[1/q - 1\right]}}{2 \pi i}\right) - \ln\left(\frac{  T_K(T_c \log{\left[1/q - 1\right]})}{2\pi T_c}\right) \right] =0,
\end{equation}
a non-linear equation for $T_c$ which has at most two solutions due to its convexity (see Appendix \ref{sec:convexity}). We solve \eqref{eq:Tcsolve} analytically for a half-filled ($q = 0.5$) impurity as a function of the $d$-level $\tilde \epsilon_d$, which simplifies to,
\begin{equation}\label{eq:Tcsolvehalffilling}
    \text{Re}\left[\psi \left(\frac{1}{2}\right) - \ln\left(\frac{  T_K(0)}{2\pi T_c}\right) \right] =0.
\end{equation}
Hence, the analytical expression for the transition temperature at half-filling is,
\begin{figure}[t]
    \centering
    \includegraphics[width=1.0\linewidth]{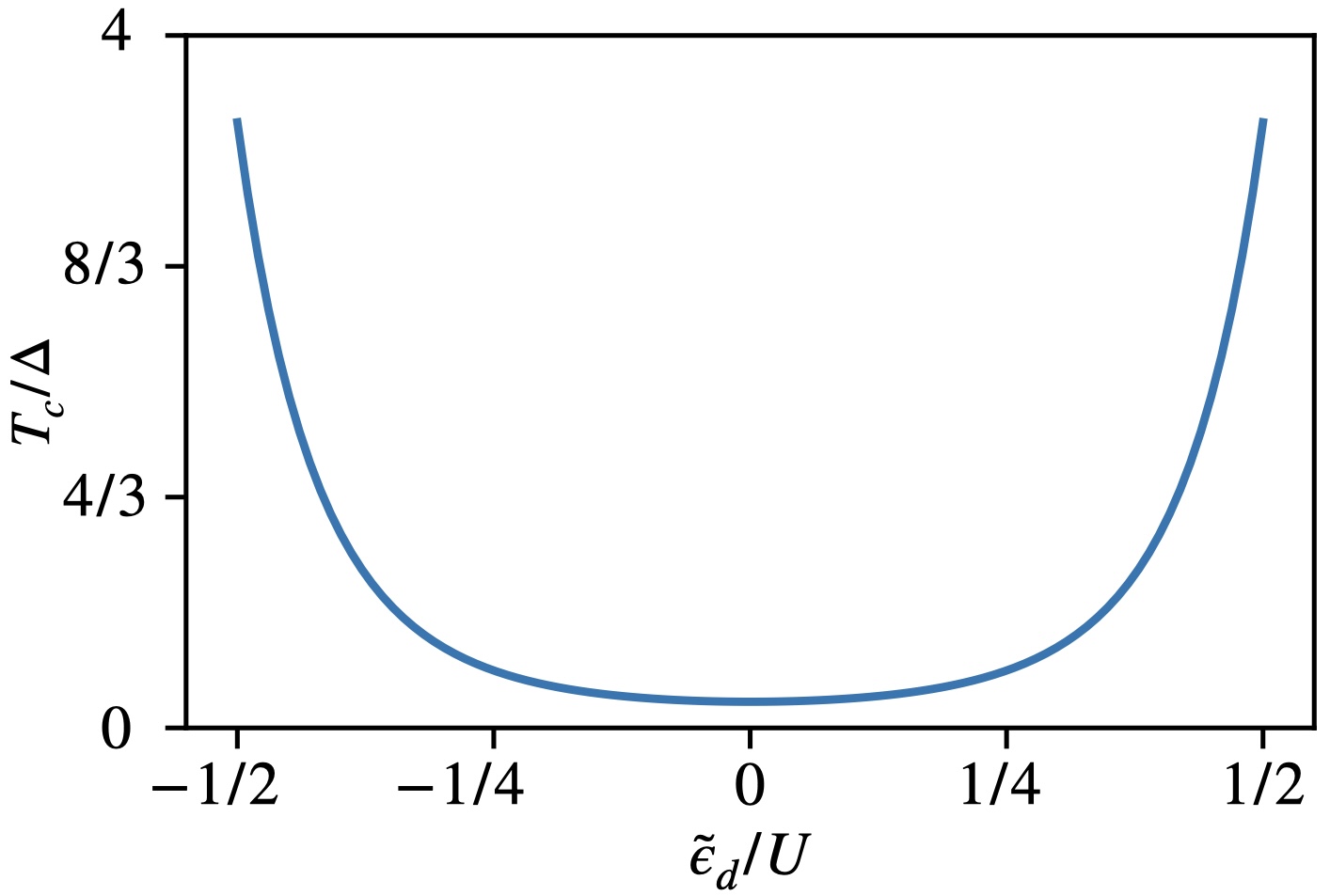}
    \caption{Mean-field transition temperature $T_c$ \lhlnew{in units of $\Delta$} as a function of $d$-level position $\tilde \epsilon_d$ for a half-filled impurity $q = Q/N = 0.5$, $\Delta = 7.5$, $U = \lhlnew{4 \Delta}$, and conduction bath \lhlnew{with bandwidth $2D$ and cutoff $D = 8\Delta / 3$}. For reference, the Kondo temperature at particle-hole symmetry $\tilde \epsilon_d = 0$ is $T_K(0, \tilde \epsilon_d = 0) = \lhlnew{0.11 \Delta }$, growing with increasing asymmetry such that the ratio of the transition temperature to the Kondo temperature $T_c^{q = 1/2}/T_K(0) =  1.13$.}
    \label{fig:Tchalffilling}
\end{figure}
\begin{equation}\label{eq:Tchalffilling}
   T_c^{q = 1/2} =  \frac{T_K(0)}{2\pi} \exp{\left\{-\psi\left(\frac{1}{2}\right)\right\}},
\end{equation}
where $T_K(0)$ is a function of $\tilde \epsilon_d$ due to the $\tilde \epsilon_d$ dependence in $J(\lambda)$ (Eq. \eqref{eq:KondoJgeneralized}). We plot the analytic mean-field transition temperature for half-filling \eqref{eq:Tchalffilling} for an Anderson impurity with a large-$N$ bare hybridization width $\Delta = \pi \rho_c V_0^2  = 7.5$, onsite $U = \lhlnew{4\Delta}$, and conduction bath \lhlnew{with bandwidth $2D$ and cutoff $D = 8\Delta/3$} (Fig. \ref{fig:Tchalffilling}).

We can also calculate the mean-field impurity valence $\langle n_f\rangle$ at zero temperature. Comparing \eqref{eq:minimizewrtlambda} and \eqref{eq:flatdosminimizewrtlambda} we obtain, 
\begin{equation}
    \langle n_f \rangle = \frac{N}{\pi}\,  \text{Im}\left[\tilde \psi (\xi) + \frac{i \pi}{2} \right].
\end{equation}
In the limit of  zero temperature, $\tilde{\psi}(\xi)$ is asymptotically $\ln{\xi/\left(2 \pi i T\right)}$, we can relate the zero temperature impurity valence $\langle n_f \rangle $ with the zero temperature value of the complex $d$-level $\xi$,
\begin{equation}\label{eq:impuritvalencezerotemp}
    \langle n_f \rangle = Q - 2\frac{\left(\tilde \epsilon_d - \lambda\right)}{U} \vert s\vert^2 = \frac{N}{\pi} \;\text{Im}\ln{\lhl{\left(\xi\right)}}.
\end{equation}
To calculate the zero temperature value of $\xi$, the complex form \eqref{comp} of the saddle-point equations is,
\begin{equation}
    \ln{\left(\frac{\xi}{T_K(\text{Re}[\xi]) \,  e^{i \pi q}}\right)} = -2 \pi i \frac{\tilde \epsilon_d - \lambda}{U} \vert \tilde s \vert^2,
\end{equation}
at zero temperature. By substituting $\lambda = \text{Re} [\xi]$, $\text{Im} [\xi] = \tilde \Delta = \pi \rho_c V_0^2 \vert \tilde s \vert^2$, and using $\Delta = \pi \rho_c V_0^2$, this equation can be rewritten in terms of the complex $d$-level position,
\begin{eqnarray}
    &&\frac{\pi}{ U \Delta  } \left[\left(\xi - \tilde \epsilon_d \right)^2 - \left(\text{Re}\left[\xi\right] - \tilde \epsilon_d \right)^2  + \text{Im}\left[\xi\right]^2\right]  \cr &=&   \ln{\left(\frac{\xi}{T_K(\text{Re}[\xi]) \,  e^{i \pi q}}\right)} ,
\end{eqnarray}
which can be solved for the zero temperature value of the complex $d$-level $\xi$. We can then immediately calculate the zero temperature mean-field impurity valence $\langle n_f \rangle$ from \eqref{eq:impuritvalencezerotemp}.

We now present the mean-field results as a function of electron filling $x$ for an $N = 6$ SIAM with a bare hybridization width in the large-$N$ analysis of $\Delta = 25$, onsite Hubbard $U = \lhlnew{4\Delta}$, and a flat conduction density of states \lhlnew{with a bandwidth $2D$ and cutoff $D = 4\Delta/5$}. 
\begin{figure}[t!]
       \centering
       \includegraphics[width=1.0\linewidth]{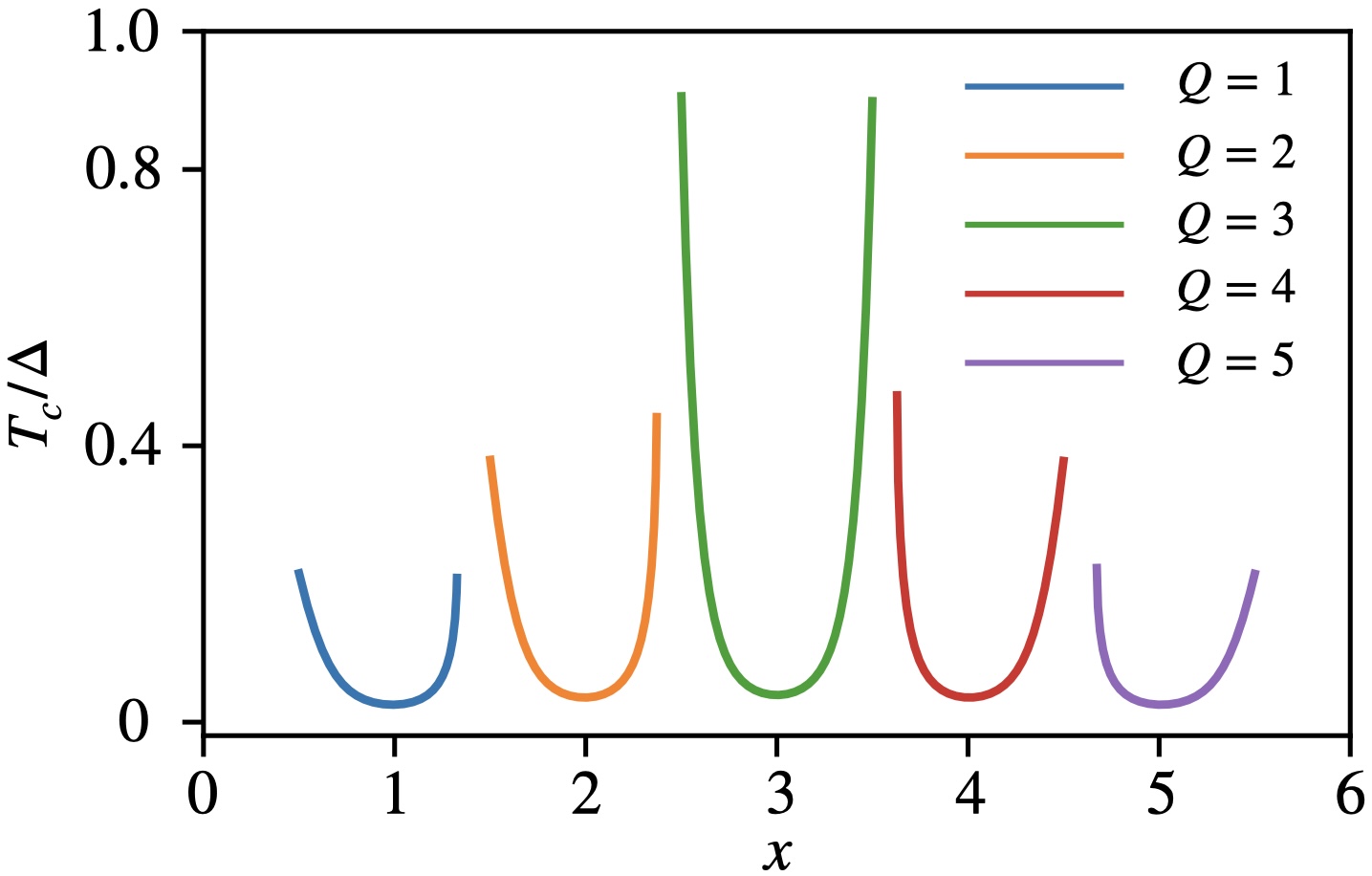}
       \caption{Mean-field transition temperature $T_c$ \lhlnew{in units of $\Delta$} plotted versus electron filling $x$ for $\Delta = 25$, $U = \lhlnew{4 \Delta} $, and a conduction bath \lhlnew{with bandwidth $2D$ and cutoff $D = 4\Delta/5 $} for different integer atomic fillings $Q$ of the $N = 6$ impurity ($Q = 1$, blue, $Q = 2$, orange, $Q = 3$, green, $Q = 4$, red, and $ Q = 5$, purple). For each $Q$, the $d$-level position from half filling $\tilde \epsilon_d$ is taken to vary linearly from $U/2$ to $-U/2$ between half-integer fillings. $x$ is restricted for each $Q$ to the range where there is at least one solution for $T_c$. We only plot the lower calculated $T_c$ solutions.}  
       \label{fig:Tcmilleniumdome}
\end{figure}

\begin{figure}[t!]
    \centering
    \includegraphics[width=1.0\linewidth]{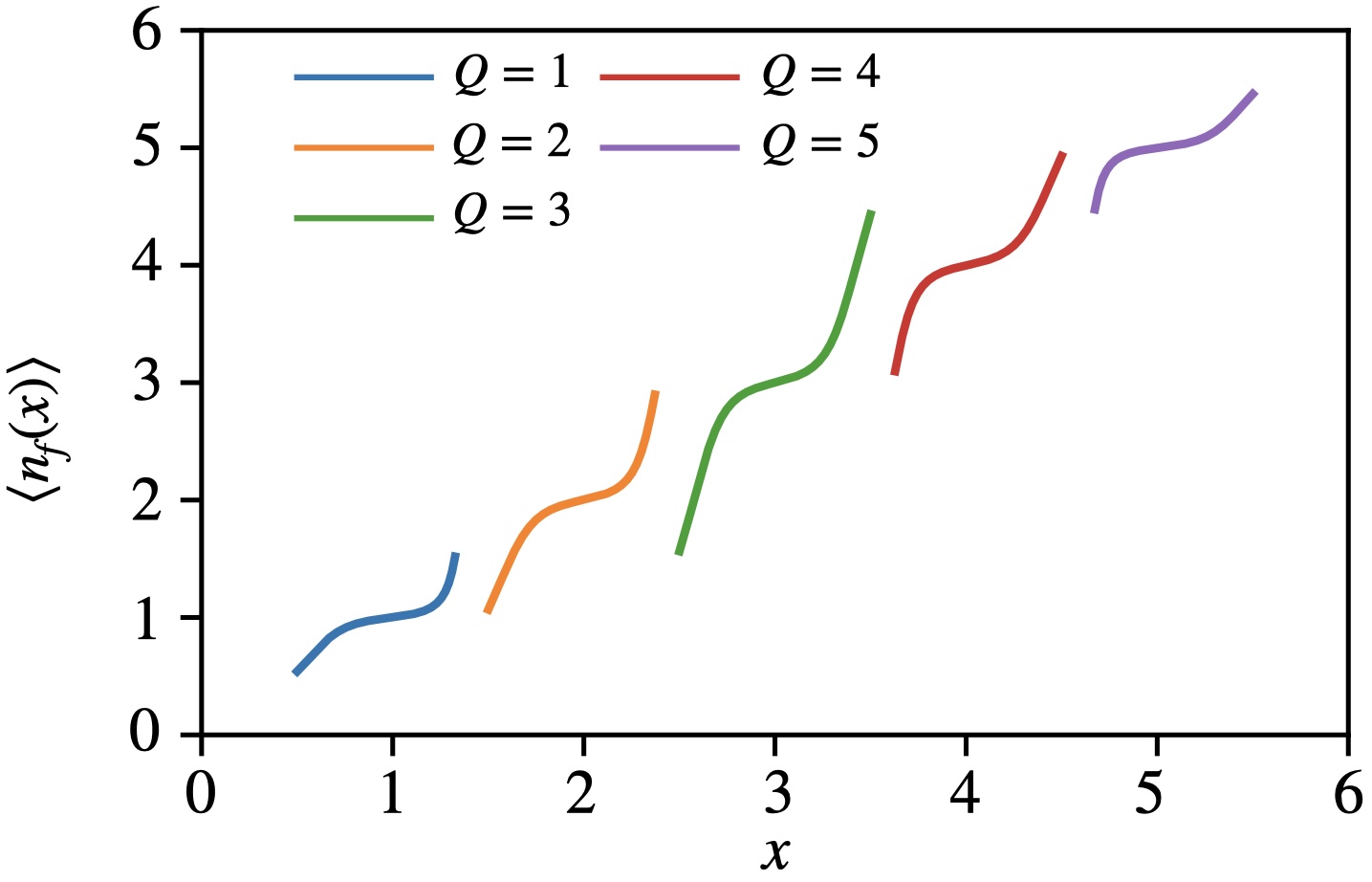}
    \caption{Mean-field impurity valence at zero temperature as a function of electron filling $x$ for $\Delta = 25$, $U = \lhlnew{4\Delta}$, and a conduction bath \lhlnew{with bandwidth $2D$ and cutoff $D = 4\Delta/5$} for different atomic fillings $Q$ of the $N=6$ impurity ($Q = 1$, blue, $Q = 2$, orange, $Q = 3$, green, $Q = 4$, red, and $ Q = 5$, purple). For each $Q$, the $d$-level position from half filling $\tilde \epsilon_d$ is taken to vary linearly from $U/2$ to $-U/2$ between half-integer fillings. $x$ is restricted for each $Q$ to the range where there is at least one solution for $T_c$.}
    \label{fig:meanfieldfilling}
\end{figure}

Following the procedure described above, we calculate the transition temperature $T_c$ into the mean-field Kondo phase as a function of electron filling $x$ (Fig. \ref{fig:Tcmilleniumdome}) , and the zero temperature impurity filling $\langle n_f\rangle$ as a function of electron filling $x$ (Fig. \ref{fig:meanfieldfilling}). We mimic the resetting of the Coulomb blockade physics (when reaching the next impurity valence) by manually controlling two parameters as a function of $x$. The electron filling $x$ is taken to be a continuous variable where $x = 0$ when the impurity is empty, and $x = 6$ is when the impurity is completely full. First, we take $Q$, the number of $d$-electrons in the atomic limit of the impurity, to increase stepwise as a function of $x$,
\begin{equation}\label{eq:mftqcontrol}
Q(x) =     \lfloor x \rceil \in \mathbf{Z}^+,
\end{equation}
and only taking positive integer values by taking the rounded value of $x$. For example, in the range $w\leq x < 2.5$, $Q = 2$, and $Q = 3$ when $2.5 \leq x <3.5$. For each integer $Q$, we produce a separate curve (either $T_c$ or $\langle n_f \rangle $) where we linearly vary the $d$-level $\tilde \epsilon_d$ and reset the value of $\tilde \epsilon_d$ for the next integer $Q$,
\begin{equation}\label{eq:mftEdcontrol}
   \tilde \epsilon_d(x) =  -U \left(x - Q(x) \right).
\end{equation}
$\tilde \epsilon_d$ equals $U/2$ at the half-integer below integer $x$, linearly decreasing to $\lhl{-}U/2$ at the half-integer above integer $x$, with $\tilde \epsilon_d = 0 $ at integer $x$. This resets for the next integer $Q$ value, running over the next $x$ range. For a given integer $Q(x)$, certain $\tilde \epsilon_d(x)$ values may not have a solution for $T_c$ (see Appendix \ref{sec:convexity}), so $x$ will be restricted for each integer $Q$ in a way to not include the values of $\tilde \epsilon_d(x)$ which have no transition temperature solution.

Generally, \eqref{eq:Tcsolve} may have two solutions, we only plot the lower transition temperature $T_c$, into the lower mean-field Kondo phase, as a function of filling $x$ with $N = 6$ in figure \ref{fig:Tcmilleniumdome} (see Appendix \ref{sec:convexity} for the same figure with the high temperature re-entrant $T_c$ included). We plot five separate curves, one for each $Q = \{1, 2,3,4,5\}$. The resulting $T_c(x)$ curves have maxima near half-integer fillings (where mixed valence is greatest) and minima at integer fillings (where the Kondo limit holds). We have restricted the $x$ ranges for Fig. \ref{fig:Tcmilleniumdome} to where at least one solution for $T_c$ exists, as discussed in the previous paragraph.

Figure \ref{fig:meanfieldfilling} shows the impurity valence $\langle n_f \rangle$ as a function of electron filling $x$, one for each $Q = \{1, 2,3,4,5\}$, for $N=6$. We follow the same procedure to restrict the $x$ range for each $Q$ curve to where at least one solution for $T_c$ exists. The value of $\langle n_f \rangle$, near integer filling $x$, is $Q(x)$ which increases in integer steps as $x$ is increased, resembling the atomic behavior. Away from integer filling, $\langle n_f \rangle$ deviates from $Q$ due to increasing mixed valence, captured in Eq. \eqref{eq:impuritvalencezerotemp}. 

The curves for $T_c$ (Fig. \ref{fig:Tcmilleniumdome}) and the zero temperature $\langle n_f \rangle$ (Fig. \ref{fig:meanfieldfilling}) do not connect smoothly or continuously for different $Q$. This is partly due the manual stepwise increase of $Q$ \eqref{eq:mftqcontrol} and reset the $d$-level $\tilde \epsilon_d$ \eqref{eq:mftEdcontrol}. Similar to the slave rotor approach \cite{physrevb.66.165111, florens_prb2004}, which exhibits reduced accuracy away from half filling ($Q = N/2$), our theoretical framework becomes unreliable when the system deviates significantly from integer occupancy values $Q$, as shown from the large deviations at strong mixed valence. \lhl{One of the possible reasons for} the large deviations (of order $N$) of the impurity valence $\langle n_f \rangle$ from its atomic value $Q$, when $x$ deviates from the integer $Q$, is the $\mathcal{O}(\sqrt{N})$ $s$-boson value.  We can estimate the impurity valence at half-integer filling $x$ where $\tilde \epsilon_d - \lambda \approx \pm U/2$, Eq. \eqref{eq:impuritvalencezerotemp} then becomes,
\begin{equation}\label{eq:strongmixedvalenceissues}
    \langle n_f \rangle \approx Q \, \mp \, \vert s\vert^2 = Q  \mp \mathcal{O}(N),
\end{equation}
agreeing with the results in Fig. \ref{fig:meanfieldfilling}. \lhl{Our method aims to capture fluctuations only to nearby valences ($Q \pm1$) in the SIAM with hybridization, as only then does it adiabatically connect to the atomic limit where the energetically relevant excitations are single electron ionizations. However, the method has difficulties in extreme mixed valence cases between integer filling factors when the valence fluctuation excitations, tracked by the $b$ and $t$ bosons, becomes energetically cheap. The constraint \eqref{eq:constraint} allows \lhlnew{both} $n_b$ and $n_t$ to become very large and no longer accurately describing the fluctuations to the nearby valence, in essence the same as what is described in \eqref{eq:strongmixedvalenceissues}. This leads to an overestimation of the degree of mixed valence away from integer fillings $Q$, which is observed in Fig. \ref{fig:meanfieldfilling}.}
In section \ref{sec:discussion}, we discuss a normalization procedure of the $s$-boson that may control our theory in the strongly mixed valent regime.

\subsection{Gaussian Fluctuations}\label{sec:Gaussianflucs}

The dynamic valence fluctuations of the SIAM can be economically captured within the method by including fluctuations around the mean-field solution \begin{equation}s(\tau) = s + \sqrt{N}\delta s(\tau)\end{equation} to second order. By expanding $S_E$ to Gaussian order, we obtain 
\begin{widetext}
\begin{eqnarray}\label{eq:gausscorrectionsF}
    \Delta S_E &=& N \sum_{i\nu_m} \bar{s}_m s_m \left\{ \frac{{1}}{U}\left[\left(\frac{U}{2}\right)^2 - \left(\lambda - \tilde{\epsilon}_d -  i \nu_m \right)^2 \right] - \chi_0(i \nu_m) \right\} - \frac{N}{2}\sum_{i \nu_m} \left[\bar{s}_m\bar{s}_{-m} + s_m s_{-m} \right] \chi_A(i\nu_m)
\end{eqnarray}
\end{widetext}
where the fluctuations of $s$ is written in Matusbara frequency $s_m \equiv\int _0^{\beta} e^{i \nu_m \tau} \delta s(\tau)$, and the hybridization susceptibilities $\chi_0(i\nu_m)$ and $\chi_A(i\nu_m)$ for a conduction bath with a constant density of states $\rho_c$ are unchanged from the infinite $U$ slave boson case \cite{coleman1987x}.

Let us consider the ``normal  state", where $\tilde s=0$ and the anomalous Bose self-energy $\Pi_A$ vanishes. The first term in the curly bracket of \eqref{eq:gausscorrectionsF} derives from the kinematics of the $s$-boson in the effective action \eqref{eq:SA2}, while the second term
is described by the one-loop Kondo susceptibility $\chi_0(i\nu_m)$ which is,
\begin{eqnarray}\label{chi0suscep}
    -\chi_0(i \nu_m) &&= \frac{\Delta}{\pi}\left(\left[\tilde{\psi}(\xi + i \nu_m) - \tilde{\psi}(\xi)\right] \right)^* + \cr &&\frac{\Delta}{\pi}\, \textrm{Re} \left[\tilde{\psi}(\xi) - \ln{\frac{D}{2 \pi i T}} \right],
\end{eqnarray}
for a conduction bath with a constant density of states (see Appendix \ref{sec:gausscartesian} for the full expression from \cite{coleman1987x}).
Here, $\Delta = \pi \rho_c V_0^2$ is the bare hybridization width.  Recall that the complex $d$-level position \eqref{cfpos}  $\xi = \lambda  + i\tilde{\Delta} = \lambda $ in the normal state,  since the imaginary part $\tilde{\Delta} = \Delta \vert \tilde s\vert^2=0$ vanishes.
We can immediately write down the inverse propagator for the $s$-boson Gaussian fluctuations in the normal state,
\begin{eqnarray}\label{eq:spropy}
    {D_s(i\nu_m)^{-1}} &=&   \chi_0(i\nu_m) - \frac{U}{4}\left[1 -  \frac{\left(\lambda - \tilde{\epsilon}_d- i \nu_m  \right)}{(U/2)^2}^2 \right]\cr &=& \langle \bar{s}_m s_m \rangle^{-1}/N .
\end{eqnarray}
\begin{figure}[t]
    \centering
    \includegraphics[width=1.0\linewidth]{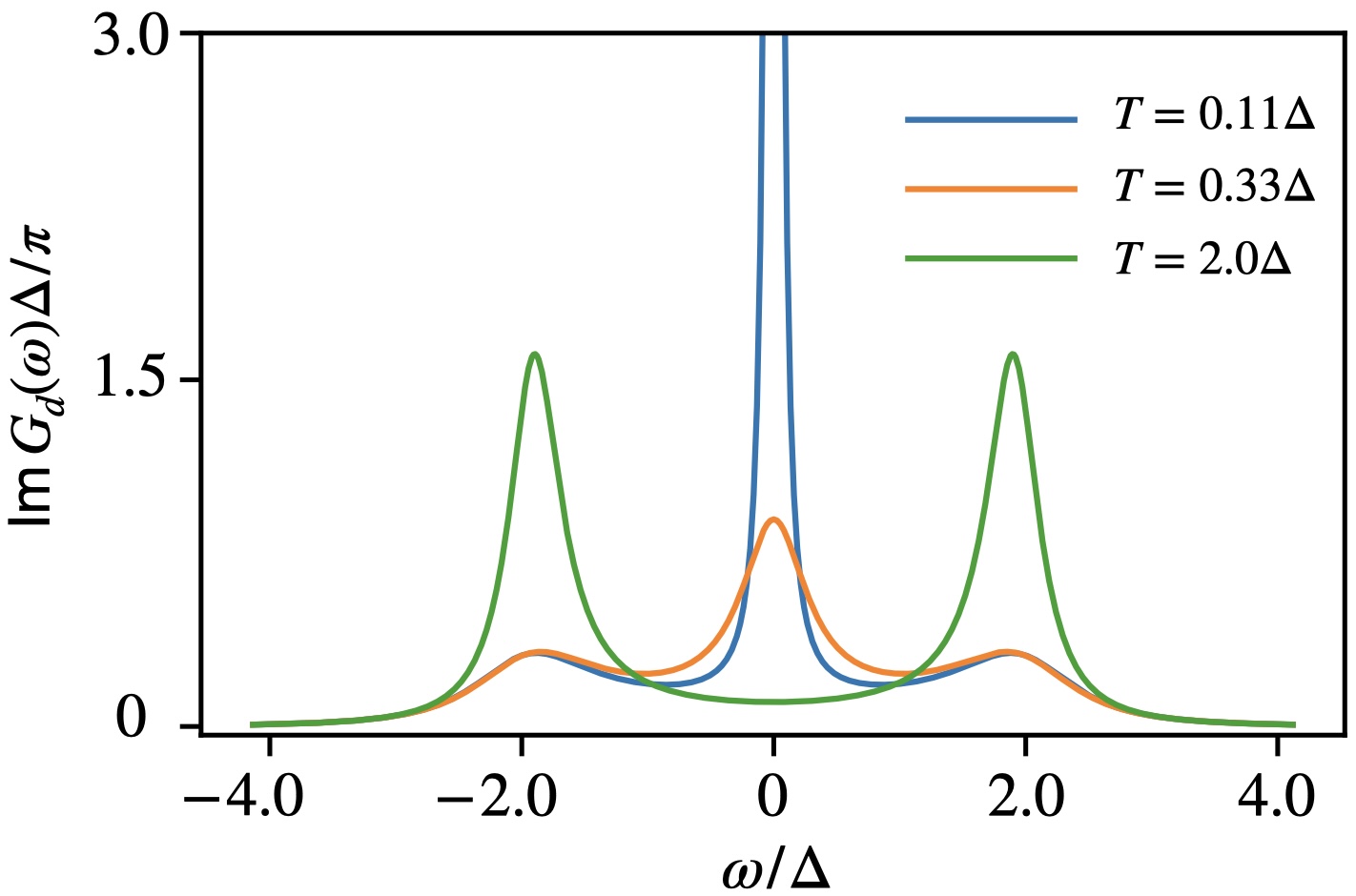}
    \caption{The full physical $d$-electron spectral function for the symmetric limit \lhlnew{(multiplied by $\Delta$ so that the y-axis is dimensionless)}, $q = 0.5$, $\Delta = 7.5$, $\tilde{\epsilon}_d = 0.0$, $U =  4 \Delta  $, and a conduction band \lhlnew{with bandwidth $2D$ and cutoff $D = 8\Delta/3$}. The Kondo mean-field transition temperature $T_c  = \lhlnew{0.13 \Delta}$ for these parameters. We plot the $d$-electron spectral function for different temperatures: high temperature ($T \gg T_c$, green); intermediate temperature ($T > T_c$, but where a central resonance begins to form, orange), and low temperature (with temperature just below $T_c$, blue).}
    \label{fig:symmetricspectral}
\end{figure}
The economy of our approach is that the atomic information in the $s$-boson propagator is fully captured by the right hand term of \eqref{eq:spropy}, whilst the low frequency term is captured by the $c$-$f$ bubble $\chi_0(i\nu_m)$, which can be cheaply calculated for other conduction density of states.
\section{Application to Single Impurity Anderson Model}\label{sec:applicationtoSIAM}
This section evaluates how accurately the $s$-boson theory with Gaussian fluctuations in the ``normal'' state reproduces the physical properties of the SIAM.
\subsection{Spectral Function Results}
The physical $d$-electron as the product of the $d$-spinon and the $s$-boson,
\begin{equation}\label{eq:physicaldelectron}
    d\dg_\sigma(\tau)= f\dg_{\sigma}(\tau) s(\tau) = f\dg_\sigma (\tau) \delta s(\tau),
\end{equation}
where we have considered the normal state.
We can then approximate the physical $d$-electron Green's function by factorizing,
\begin{eqnarray}\label{eq:delectrongreens}
    G_d(\tau) &=& -\langle \mathcal{T} d_\sigma (\tau) d\dg_\sigma(0)\rangle = -\langle \mathcal{T} \delta s\dg(\tau) f_\sigma (\tau) f\dg_\sigma(0) \delta s(0) \rangle \cr &\approx&-\langle \mathcal{T}  f_\sigma (\tau) f\dg_\sigma(0) \rangle \langle \delta s\dg(\tau) \delta s(0) \rangle \cr &=& - G_{f}(\tau) D_s(-\tau).
\end{eqnarray}
where $G_f(\tau)$ and $D_s (\tau)$ are the $f$-spinon and $s$-boson Green's functions, respectively. Using the convolution theorem, 
\begin{equation}\label{eq:bosonicmatsubarasum}
    G_d(i\omega_n) = -\frac{1}{\beta} \sum_{i\nu_r} G_{f}(i\omega_n + i\nu_r)D_s(i\nu_r),
\end{equation}
where the $f$-spinon propagator is,
\begin{equation}
    G_f (i \omega_n) = \left(i\omega_n - \lambda\right)^{-1}.
\end{equation}
Performing the bosonic \lhl{M}atsubara sum \label{eq:bosonicmatsubarasum} as an integral on the complex plane,
\begin{eqnarray}
    G_d(i\omega_n) &=&   \ointclockwise_C \frac{dz \; n_B(z)}{2 \pi i} G_f(i \omega_n + z) D_s(z) \cr &=& \int \frac{dv}{\pi}D^{''}_s(v - i\delta) \left[\frac{n_B(v) +q}{i \omega_n -\lambda + v}\right],
\end{eqnarray}
where the contour $C$ is taken to be clockwise around the poles and branch cuts of $G_f$ and $D_s$. Here $n_B(\omega) = 1/ \left(e^{\beta \omega} - 1\right)$ is the Bose-Einstein distribution and we have used $n_B(\lambda - i \omega_n) = - f(\lambda) = \lhl{-}q$, where $f(\lambda)= 1/ (e^{\beta \lambda} +1) $ is the Fermi function and \lhl{$q = Q/N$ is the ratio of the integer number of $d$-electrons and the total number of $d$-electron flavors}.
The imaginary part of the $d$-electron propagator is then related to the spectral function,
\begin{eqnarray}\label{eq:spectralfunc}
    A_d(\omega) &=& G^{''}_d(\omega - i\delta)/\pi \cr &=& \frac{1}{ \pi}\left[n_B(\lambda  - \omega ) + q \right]\,  D^{''}_s(\lambda - \omega),
\end{eqnarray}
which can be calculated from the $s$-boson propagator at the level of Gaussian fluctuations.
\begin{figure}[t]
    \centering
    \includegraphics[width=1.0\linewidth]{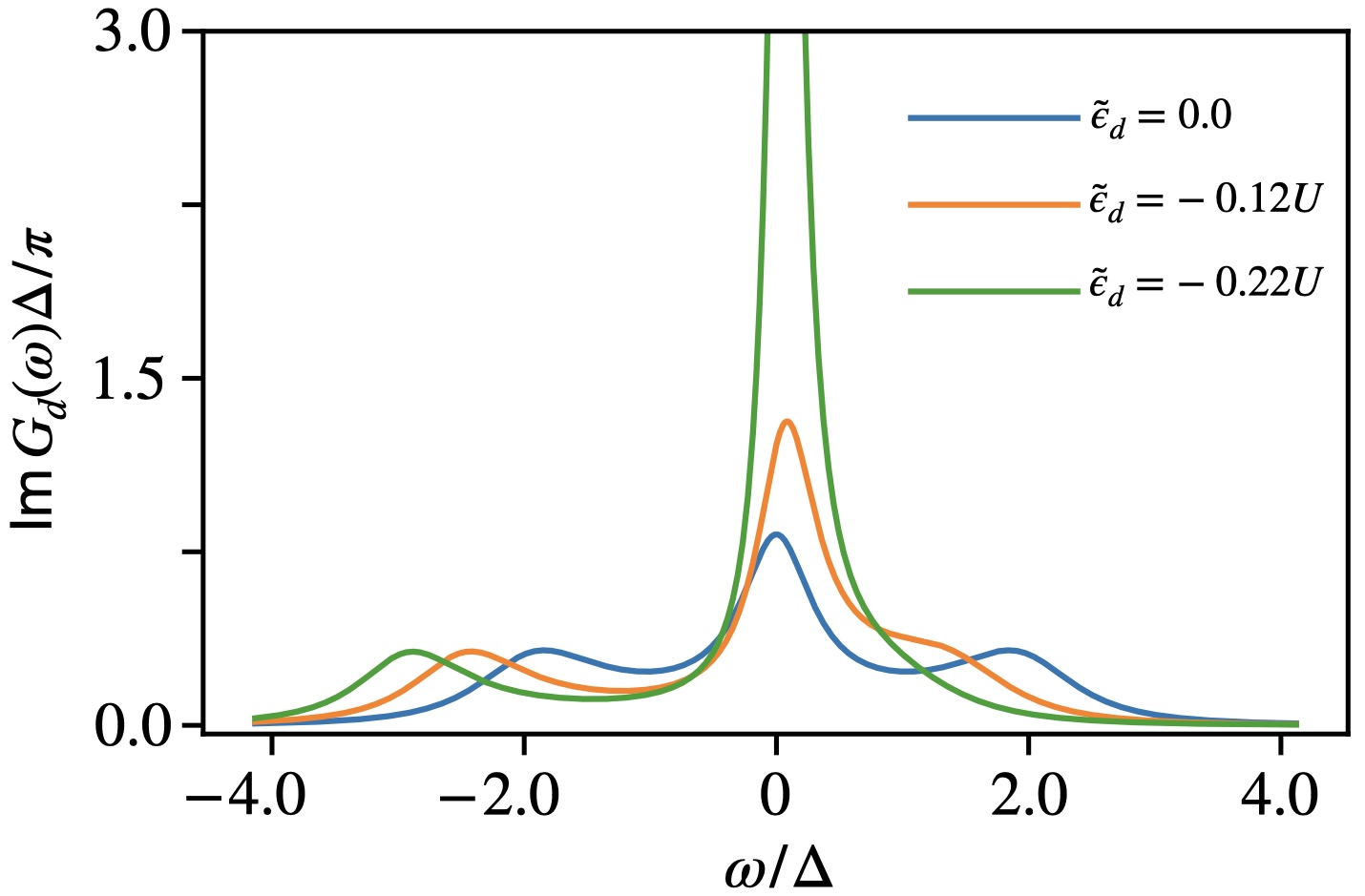}
    \caption{The $d$-electron spectral function for the asymmetric limit \lhlnew{(multiplied by $\Delta$ so that the y-axis is dimensionless)}, $q = 0.\lhl{5}$, $\Delta  = 7.5$, $T = \lhlnew{0.35 \Delta}$, and $U = 4\Delta$, and a conduction band \lhlnew{with bandwidth $2D$ and cutoff $D = 8\Delta/3$} for different $d$-electron levels. The temperature $T$ is chosen to be greater than the transition temperatures $T_c$ for the plotted parameters.}
    \label{fig:asymmetricspectral}
\end{figure}

For the normal state, we take the $d$-level to be determined by the thermal occupancy of the $d$-electron $\lambda= T \ln{\left[\left(1/q\right) -1\right]}$. In Fig. \ref{fig:symmetricspectral}, we plot the spectral function of the full physical $d$-electron in the particle-hole symmetric limit $q = 0.5$, and $\tilde \epsilon_d = 0$ with the large-$N$ bare hybridization width $\Delta = 7.5$, onsite interaction $U = \lhlnew{4\Delta}$, and a conduction band \lhlnew{with bandwidth $2D$ and cutoff $D = 8\Delta/3$}. 

We can use these parameters and analytically calculate \lhlnew{a Kondo temperature $T_K = \lhlnew{0.11\Delta} $ \eqref{eq:TKlargeN} and then get the transition temperature $T_c/T_K = 1.13$ \eqref{eq:Tchalffilling} from the analytic result at half-filling}. At high temperatures ($T \gg T_c$, there is an upper Hubbard band at $\omega = U/2$ and lower Hubbard band at $\omega = -U/2$ (see the green curve of Fig. \ref{fig:symmetricspectral}). Interestingly, even within the normal phase of the $s$-boson when it has yet to condense and the anomalous self energy $\Pi_A(i\nu_m)$ is still zero, a central Kondo resonance begins to develop as the temperature is lowered, but still above the transition temperature $T_c$ (see the orange curve of Fig. \ref{fig:symmetricspectral}). We plot the results for the high temperature limit ($T \gg T_c$), intermediate temperature ($T > T_c$) where a central resonance begins to form, and a low temperature result just below $T_c$.

In fig. \ref{fig:asymmetricspectral}, we plot the spectral function of the full physical $d$-electron for the Anderson impurity model as it goes from being symmetric to becoming more and more asymmetric by tuning the $\tilde{\epsilon}_d$ level for a fixed temperature $T$ that is greater than the mean-field transition temperature $T_c$ for each parameter regime. We use the parameters $q = 0.5$, with the large-$N$ bare hybridization width $\Delta = 7.5$, onsite interaction $U = \lhlnew{4 \Delta}$, and a conduction band \lhlnew{with bandwidth $2D$ and cutoff $D = 8\Delta/3$}. The temperature is kept in the normal state to ensure physically meaningful results, when limited to our normal state analysis, for the dynamical charge susceptibility results presented in the next subsection.

\subsection{Charge Susceptibility}\label{sec:chargesusceptibility}

The dynamic $f$-spinon density can be calculated by differentiating the action \eqref{eq:Csummary} with respect to $\lambda$, we obtain the following Ward identity,
\begin{equation}\label{eq:wardidentity}
 n_f(\tau)  = Q + \frac{2}{U} \bar{s}(\tau) \left(\lambda - \tilde \epsilon_d+ \overleftrightarrow{\partial_\tau}\right)s(\tau),
\end{equation}
where $\bar{s} \overleftrightarrow{\partial_\tau} s \equiv \frac{1}{2} \left( \bar{s}\, \partial_\tau[s]  - \partial_{\tau} [\bar{s}] s \right)$. Taking the fluctuations of both sides of \eqref{eq:wardidentity}, we can relate the fluctuations of $n_d$ with the Gaussian fluctuations of $s(\tau)$,
\begin{equation}\label{eq:deltanftau}
    \delta n_f(\tau) = \frac{2N}{U} \delta \bar{s} \left(\lambda - \tilde \epsilon_d+ \overleftrightarrow{\partial_\tau}\right) \delta s.
\end{equation}
The imaginary time charge susceptibility is defined as,
\begin{equation}\label{eq:imaginarytimechargesusc}
   \chi_c(\tau) = \langle \mathcal{T} \delta n_f(\tau) \delta n_f (0) \rangle.
\end{equation}
The Fourier transform of this yields the \lhl{M}atsubara frequency dependent charge susceptibility $\chi_c(i \nu_m) = \langle \delta n_f (-i\nu_m) \delta n_f (i\nu_m) \rangle$. By Fourier transforming \eqref{eq:deltanftau} and comparing $\chi_c(i \nu_m)$ with the $s$-boson propagator \eqref{eq:spropy}, the \lhl{M}atsubara frequency dependent charge susceptibility can be related to the $s$-boson bubble with the vertex $v_l = 2(\lambda - \tilde \epsilon_d -i\omega_l -i\nu_m/2)/U $, coming from the parenthesis in \eqref{eq:deltanftau},
\begin{eqnarray}\label{eq:chargesusceptibility}
    \chi_c(i\nu_m) &=& \frac{4  T}{U^2}\sum_{i\omega_l}D_s(i\nu_m+i\omega_l) D_s(i\omega_l)\cr &\times& \left[\lambda - \tilde \epsilon_d - i\omega_l -i\nu_m/2\right]^2.
\end{eqnarray}
Analytically continuing to the real frequency $\nu$ axis, 
\begin{eqnarray}\label{eq:chargesusceptibilityrealfreq}
    \chi_c(\nu - i\delta) &=& \frac{4}{U^2}\int_{-\infty}^{\infty}\frac{d \omega}{\pi}\int_{-\infty}^{\infty}\frac{d \omega'} {\pi}  \frac{n_B(\omega) - n_B(\omega')}{\nu - i\delta + \omega - \omega'} \cr &\times& D^{''}_s(\omega')\,D^{''}_s(\omega)\left[\lambda - \tilde\epsilon_d -\left(\frac{\omega'}{2}+ \frac{\omega}{2} \right)  \right]^2.
\end{eqnarray}
So the imaginary part of the charge susceptibility is,
\begin{eqnarray}\label{eq:dynamicalchargesusceptibility}
    \chi^{''}_c(\nu) &=& \frac{4}{U^2}\int_{-\infty}^{\infty} \frac{d\omega}{\pi} \left(n_B(\omega) - n_B(\omega+\nu)\right) D^{''}_s(\nu+\omega)\cr  &\times&D^{''}_s(\omega)\left[-\left(\frac{\nu}{2}+ \omega \right) + \Lambda \right]^2 .
\end{eqnarray}

\begin{figure}[t]
    \centering
    \includegraphics[width=1.0\linewidth]{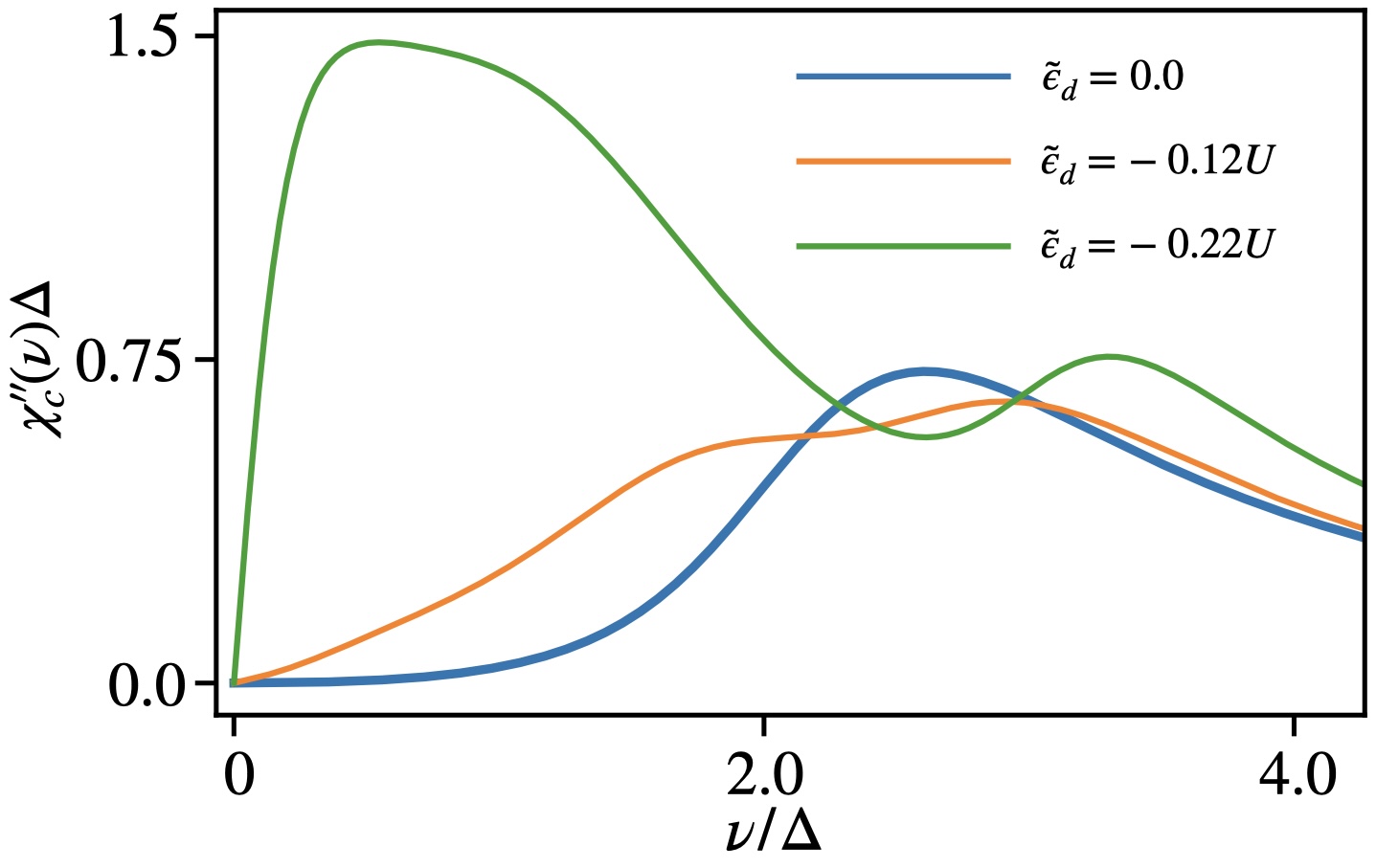}
    \caption{The imaginary part of the dynamical charge susceptibility \lhlnew{(multiplied by $\Delta$ so that the y-axis is dimensionless)} for $q = 0.5$, $\Delta= 7.5$, $T = \lhlnew{0.35 \Delta}$, and $U = 4\Delta$, and a conduction band cutoff $D = \lhlnew{8\Delta/3}$ for different $d$-electron levels $\tilde \epsilon_d$. The calculated Kondo mean-field transition temperature $T_c = \lhlnew{0.13 \Delta}$.}
    \label{fig:chargesusc}
\end{figure}
In Fig. \ref{fig:chargesusc}, we plot the imaginary part of the dynamical charge susceptibility $\chi_c^{''} (\nu)$ for the same parameters as Fig. \ref{fig:asymmetricspectral}. In the symmetric limit when $\tilde \epsilon_d = 0$, there is a single peak at the atomic excitation energy $U/2$, corresponding to both the upper and lower Hubbard band positions for the symmetric limit $\tilde \epsilon_d = 0$ in Fig. \ref{fig:asymmetricspectral}, which then splits into two peaks, at the two atomic excitation energies \eqref{eq:ionizationenergies}, separated by $2 \vert \tilde \epsilon_d \vert$. As the upper Hubbard band merges into the central resonance (Fig. \ref{fig:asymmetricspectral}), the valence fluctuations become low frequency and the dynamical charge susceptibility $\chi_c^{''}(\nu)$ develops low frequency support.

\begin{figure}[t]
    \centering
    \includegraphics[width=1.0\linewidth]{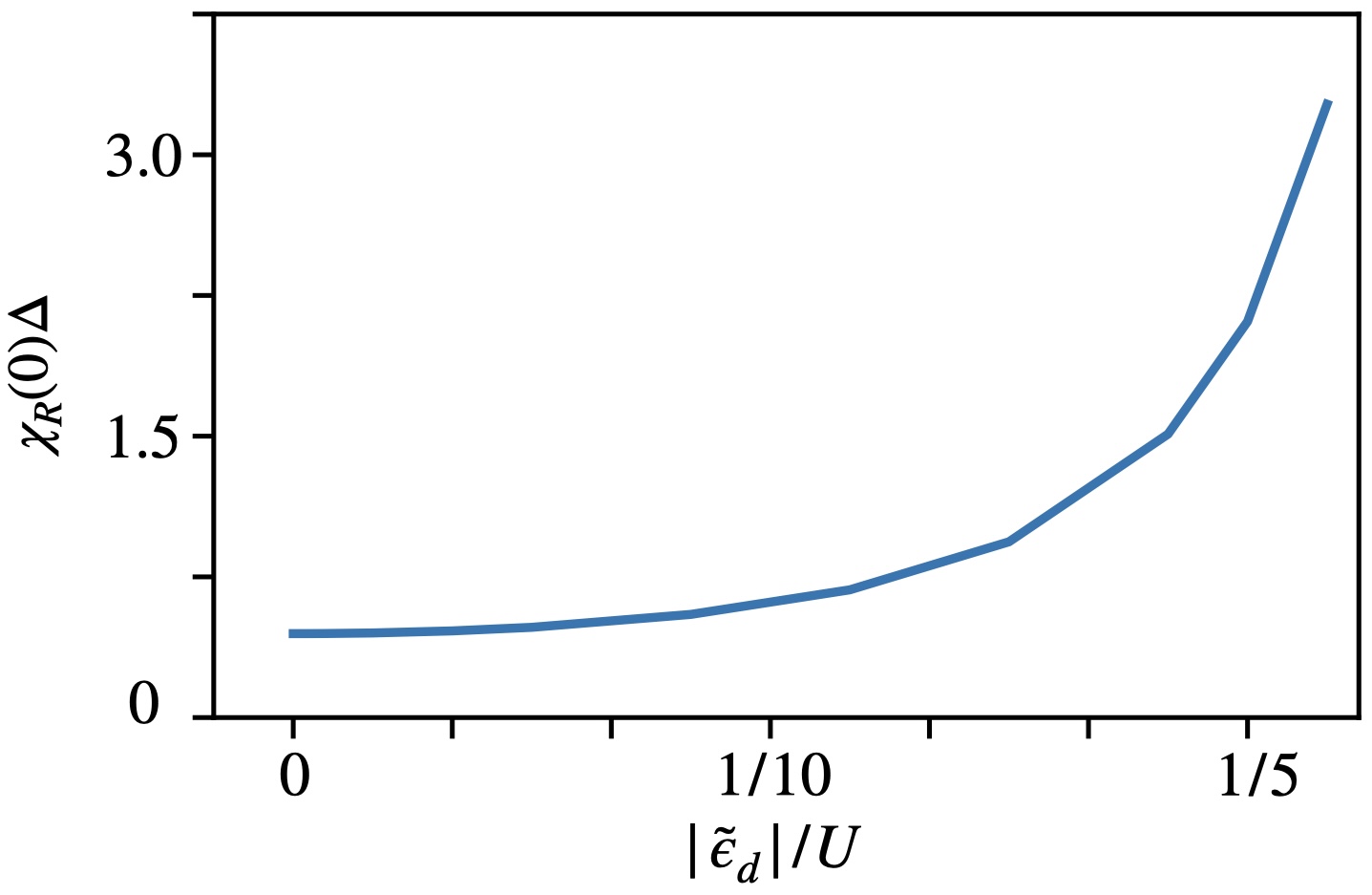}
    \caption{The static charge susceptibility $\chi_R(0)$ \lhlnew{(multiplied by $\Delta$ so that the y-axis is dimensionless)} for $q = 0.5$, $\Delta= 7.5$, $T = \lhlnew{0.35 \Delta}$, and $U = 4\Delta$, and a conduction band cutoff $D = \lhlnew{8\Delta/3}$ as a function of $d$-electron level $\tilde \epsilon_d$.}
    \label{fig:staticchargesusc}
\end{figure}

We can also calculate the static charge susceptibility via the retarded real frequency charge susceptibility,
\begin{equation}
    \chi_R(\omega + i\delta) = \int_{-\infty}^{\infty} \frac{d \nu}{\pi} \frac{\chi^{''}_c(\nu + i\delta)}{\nu - (\omega + i\delta)},
\end{equation}
taking the static limit,
\begin{equation}
    \chi_R(0) = \int_{-\infty}^{\infty} \frac{d \nu}{\pi} \frac{\chi^{''}_c(\nu + i\delta)}{\nu}.
\end{equation}
We have calculated this integral  numerically. Fig. \ref{fig:staticchargesusc} shows the resulting static charge susceptibility $\chi_R(0)$ as a function of $d$-electron level $\tilde \epsilon_d$, for $q = 0.5$, $\Delta= 7.5$, $T = \lhlnew{0.35\Delta}$, and $U = 4\Delta  $, and a conduction band cutoff $D = \lhlnew{8\Delta/3}$ (the same parameters for Figs. \ref{fig:asymmetricspectral} and \ref{fig:chargesusc}).

We find that due to the increased valence fluctuations at low frequency, the static charge susceptibility increases as the asymmetry (controlled by the absolute value of $\tilde \epsilon_d$) increases, causing the upper or lower band in the spectral function to approach zero, giving rise to a strongly mixed-valent regime.


\section{Discussion}\label{sec:discussion}

In this paper, we have have introduced a symmetric  ``$s$''-boson approach that captures valence fluctuations that both increment and decrement the ionic charge of the finite $U$ single impurity Anderson model.  By integrating out the asymmetric boson dynamics, our theory reduces to a \textit{single} symmetric boson (``$s$-boson") theory in which the upper and lower Hubbard bands are encoded in the boson kinetics. 
Our calculations of  the local spectral function and dynamical charge susceptibility  using our method demonstrate that the Kondo resonance in the SIAM begins to form at temperatures substantially above the 
 $s$-boson condensation point. This may be related to the observation from  ARPES spectroscopy, that f-bands in heavy fermion materials begin to form at temperatures considerably  above the coherence temperature as measured by  transport and thermodynamics,  for example in the 115 compound $\rm{CeCoIn}_5$ \cite{coherencetemperature}. 

The $s$-boson approach provides an efficient upgrade to the original infinite $U$ slave boson method, replacing a single pole excitation at $\omega = \lambda$ by two excitations of the $s$-boson at  $\omega = \lambda \pm U/2$, while preserving the same fermionic RPA bubbles that encode the low energy Kondo resonance information. The $s$-boson representation correctly recovers both the infinite $U$ and Kondo limits. 

One of the difficulties presented by our treatment, is its inability to handle regions of strong mixed valence character, as evidenced by  the discontinuous evolution between different $T_c$ curves for each atomic impurity filling $Q$ at half-integer filling, where mixed valence is the strongest (Fig. \ref{fig:Tcmilleniumdome}). These limitations arise from the enlarged Hilbert space (Fig. \ref{fig:fig1}) inherent in our parton formulation. While fictitious states with a large number of $b$ and $t$ slave bosons are energetically suppressed by the large atomic excitation energies in the Kondo limit, the lack of fidelity between the s-boson and the original Anderson model becomes  problematic in the strongly mixed valent regime. As either charge gap $\Delta E_\pm$ approach zero, these fictitious states become degenerate with physical charge states, allowing the $t$ and $b$ slave boson occupation numbers to diverge. This difficulty is also seen in the zero temperature impurity valence (Fig. \ref{fig:meanfieldfilling}) around half-integer filling where deviations from the atomic limit filling $Q$ becomes $\mathcal{O}(N)$ and therefore substantial and inaccurate.

\subsection{Link between the s-boson and the slave rotor}
\lhl{
We now discuss the  potential for further improvements in our methodology by examining the relationship between the s-boson and slave rotor approaches, which as we now show share a common
algebraic structure and path-integral representation.}

\lhl{Slave rotors \cite{physrevb.66.165111, florens_prb2004}, employ the phase $\theta$, dual to the impurity charge as the collective variable to capture valence fluctuations,
\begin{equation}\label{eq:originalslaverotor}
    d\dg_\sigma \rightarrow f\dg_\sigma e^{i\theta},
\end{equation}
with the constraint, 
\begin{equation}\label{eq:rotorconstraint}
    \frac{N}{2} = n_f - L_z,
\end{equation}
where the rotor operators are expressed in terms of angular momentum operators, identifying  $L_z= -i d/d\theta$, and  $L_{\pm}= e^{\pm i\theta}$. This decoupling \eqref{eq:originalslaverotor} can be regarded as a unitary transformation $d\dg_{\sigma} = e^{- i \theta n_f}f\dg_{\sigma}e^{i \theta n_f}$, and as such, it preserves the canonical commutation relations.}

\lhl{The rotor operators of the slave rotor in terms of angular momentum operators satisfy the commutation algebra $[L_z, L_{\pm} ] = \pm L_{\pm}$ and $[L_+,L_-]=0$. Generalizing the rotor constraint \eqref{eq:rotorconstraint} from being around half-filling ($Q = N/2$) to arbitrary integer filling $Q$, $n_f = Q-L_z$, analogous to the $s$-boson constraint $n_f = Q- (t\dg t - b\dg b)$. This suggests the identification $L_z \sim t\dg t - b\dg b$. The $s$-boson and its two slave boson components satisfy $[t\dg t - b\dg b, s]=s$,   $[t\dg t - b\dg b, s\dg]=-s\dg$ and $[s,s\dg]=0$,  confirming $(t\dg t - b\dg b,s) \sim (L_z,L_+)$ share identical algebras. Thus, the $s$-boson provides a faithful representation of the $U(1)$ rotation group with $s \propto L_+$.}


The key distinction between representations lies in the normalization of the algebras, which is inherently linked to whether a controlled large-$N$ saddle-point approximation is present. In our $s$-boson representation of the SIAM, there is a controlled large-$N$ saddle-point approximation in the condensed phase because $\bar s s = \vert s \vert^2 \sim \mathcal{O}(N)$, whereas the most basic $\mathcal{O}(2)$ rotor does not because the modulus of the rotor bosonic field is always of order unity. This difference manifests in a generalized anticommutation relation for the $s$-boson representation of the physical $d$-electron\eqref{eq:twobosonrep}, 
\begin{equation}\label{eq:twobosonrepanticommutation}
    \{d_\sigma, d\dg_{\sigma'}\} = \{f_\sigma s\dg, f\dg_{\sigma'} s\} =   s\dg s \,\delta_{\sigma \sigma'} = \vert s \vert^2 \delta_{\sigma \sigma'},
\end{equation}
which is only faithful when large-$N$ is abandoned and $\vert s\vert^2 = 1$. Now since $s\dg s$ commutes with $s$,  we may normalize the
algebra by writing 
\begin{equation}\label{normy}
    d\dg_{\sigma} = f\dg_{\sigma} \frac{s}{\sqrt{s\dg s}}.
\end{equation}
This restores the normalization of the canonical commutation relation $\{d_\sigma, d\dg_{\sigma'}\} = \delta_{\sigma\sigma'}$.  By comparing \eqref{normy} with \eqref{eq:originalslaverotor}, we  
identify the slave rotor phase with the phase of the $s$-boson:
\begin{equation}\label{rel}
s=\sqrt{s\dg s} e^{i \theta}
\end{equation}
yielding the fundamental relation
\begin{eqnarray}
 e^{i \theta} &=& L_+=   \frac{s}{\sqrt{s\dg s}}(1-P_{s=0}),\cr
 s &=& b + t\dg.
 \end{eqnarray}
 where $P_{s=0}$ projects out the states with $s=0$. 

Both representations successfully capture the lower and upper \lhl{Hubbard} bands and central Kondo resonance features in the SIAM. Taking a cue from these comparisons, we now demonstrate that when a constant magnitude constraint is enforced, the $s$-boson and slave rotor approaches share a common action.  Suppose we are willing to forgo a controlled large $N$ expansion, implementing the constraint $s\dg s = s_0^2 $, by including an additional term $\frac{U (h - 1)}{4} \left(\bar s s - s_0^2\right)$ in the Lagrangian \eqref{eq:Csummary}, then taking account of the normalization of $s$ to reproduce an Anderson model with hybridization $V$, we must write
\begin{widetext}
\begin{eqnarray}\label{eq:bosonactionnormalized}
    S&=& S_c+ \int_0^{\beta}d\tau \left[
     \frac{U}{4}\bar{s}
   \biggl(1- \frac{(\partial_{\tau}  + \lambda -\tilde{\epsilon}_d )^2}{(U/2)^2} \biggr) s +\sum_\sigma f\dg_{\sigma} (\partial_{\tau} + \lambda) f_{\sigma} +\frac{V}{s_0}\sum_{\bk \sigma} \left(c\dg_{\bk \sigma} f_{\sigma}{\color{black} \bar s}   + \text{H.c.} \right) +\frac{U\left(h - 1\right)}{4}  \left( \bar{s s} -s_0^2\right) - \lambda Q\right] \cr &=&S_c+ \int_0^{\beta}d\tau \left[
     \frac{U}{4}\bar{s}
   \biggl(h- \frac{(\partial_{\tau}  + \lambda -\tilde{\epsilon}_d )^2}{(U/2)^2} \biggr) s +\sum_\sigma f\dg_{\sigma} (\partial_{\tau} + \lambda) f_{\sigma} +\frac{V}{s_0}\sum_{\bk \sigma} \left(c\dg_{\bk \sigma} f_{\sigma}{\color{black}\bar s}   + \text{H.c.} \right) -\frac{Uhs_0^2}{4}   - \lambda Q\right],
\end{eqnarray}
where $S_c$ is the conduction electron action that is bilinear in the Grassman $c$.

The $O(2)$ rotor action from Florens and Georges \cite{physrevb.66.165111}, $S_{\text{rotor}} = \int_0^{\beta} d \tau \, L_{\text{rotor}}$, can be written by writing the slave rotor field $e^{i \theta}\rightarrow X$ as a complex field $X$,  subject to the constraint $|X|^2=1$. The rotor  is then,
\begin{eqnarray}\label{eq:rotorlagrange}
   &L_{\text{rotor}}& = \sum _{\bk, \sigma} c\dg_{\bk \sigma }(\partial_{\tau}+ \epsilon_{\bk})c_{\bk \sigma} +  \sum_\sigma f\dg_\sigma \left(\partial_\tau + \tilde \epsilon_d - \lambda\right) f_\sigma + \tilde h \left(\vert X\vert^2 - 1\right) \cr &+& \lambda Q - \frac{\lambda^2}{2U} \vert X \vert ^2 +  \frac{\vert \partial_\tau X\vert ^2}{2U} + \frac{\lambda}{2U}\left(\bar X \partial_\tau X - \text{h.c.}\right)  + V\sum_{\bk \sigma} \left(c\dg_{\bk \sigma} f_{\sigma}{\color{black}\bar X} + \text{H.c.} \right),
\end{eqnarray}
where we have generalized the constraint to arbitrary filling $Q$ rather than half-filling $Q = N/2$. Here, the field $\tilde h$ imposes the constraint $|X|^2=1$. The terms bilinear in $X$ fields are not $\mathcal{O}(N)$ and thus the slave rotor does not have a controlled large-$N$ expansion.

We will now transform rotor Lagrangian to show its common structure with the s-boson. First, we relabel $\lambda \rightarrow -\lambda$, and shift $\lambda \rightarrow \lambda - \tilde{\epsilon}_d$. Using the following two identities which discard nondynamical constants,
\begin{equation}
    \nonumber -\int_0^{\beta}d\tau\; \bar{X}\partial_\tau^2X = \int_0^{\beta}d\tau\; \vert\partial_\tau X\vert^2, \quad \int_0^{\beta}d\tau\; \bar{X}\partial_\tau X = -\int_0^{\beta}d\tau\;\partial_\tau[\bar{X}]X,
\end{equation}
\eqref{eq:rotorlagrange} can be rewritten as,
\begin{eqnarray}\label{eq:rotorlagrangeappx}
   &L_{\text{rotor}}& = L_c +  \sum_\sigma f\dg_\sigma \left(\partial_\tau + \lambda\right) f_\sigma + \tilde h \left(\vert X\vert^2 - 1\right)  + V\sum_{\bk \sigma} \left(c\dg_{\bk \sigma} f_{\sigma} {\color{black} \bar{X}} + \text{H.c.} \right)\cr &-&\frac{ \left(\lambda - \tilde{\epsilon}_d\right)N}{2} - \frac{\left(\lambda-\tilde{\epsilon}_d\right)^2}{2U} \vert X \vert ^2 -  \frac{\bar X \partial_\tau^2 X}{2U} - \frac{ \left(\lambda - \tilde{\epsilon}_d\right)}{U} \bar X \partial_\tau X  ,
\end{eqnarray}
where $L_c = \sum _{\bk, \sigma} c\dg_{\bk \sigma }(\partial_{\tau}+ \epsilon_{\bk})c_{\bk \sigma}$. Factoring terms, the rotor action is now,
\begin{eqnarray}\label{eq:rotoraction}
    S_{\text{rotor}} =  \int_0^\beta d\tau \left[\sum_\sigma f\dg_\sigma \left(\partial_\tau + \lambda \right) f_\sigma  + \frac{U}{8} \bar{X} \left(\frac{8}{U} \tilde h- \frac{\left( \partial_\tau + \lambda - \tilde{\epsilon}_d\right)^2}{(U/2)^2}\right)X + V \sum_{\bk \sigma} \left(c\dg_{\bk \sigma} f_{\sigma} \bar{X} + \text{H.c.} \right)- \tilde h - \lambda Q\right].
\end{eqnarray}
If we rescale $ X /\sqrt{2} \rightarrow X $,  and $h = 8 \tilde h/U$, then the rotor action becomes 
\begin{eqnarray}\label{eq:rotoractiontwo}
    S_{\text{rotor}} =  \int_0^\beta d\tau \left[\sum_\sigma f\dg_\sigma \left(\partial_\tau + \lambda \right) f_\sigma  + \frac{U}{4} \bar{X} \left(h- \frac{\left( \partial_\tau + \lambda - \tilde{\epsilon}_d\right)^2}{(U/2)^2}\right)X + \sqrt{2} V \sum_{\bk \sigma} \left(c\dg_{\bk \sigma} f_{\sigma} \bar{X} + \text{H.c.} \right)- \frac{U h}{8} - \lambda Q\right].
\end{eqnarray}
Remarkably, our compact forms for the  rotor and the s-boson action, \eqref{eq:rotoractiontwo} and \eqref{eq:bosonactionnormalized} respectively, are equivalent with  the normalization choice $s\dg s = s_0 ^2 = \frac{1}{2}$, establishing a precise equivalence between the two approaches. 
\end{widetext}

\subsection{Future directions}


Several extensions of our approach warrant future investigation. 
The identification of the slave rotor and the normalized $s$-boson SIAM actions clearly allows for future methodological advances. Writing the rotor action in a simpler $s$-boson form \eqref{eq:rotoractiontwo} allows for Gaussian fluctuations, akin to what was presented in Sec. \ref{sec:Gaussianflucs}, to be calculated simply, adjusting $h(T)$ as a function of temperature $T$ to maintain the normalization of $\langle s\dg s\rangle = s_0^2$ calculated from the Gaussian fluctuations.  If, as appears likely that the difficulties of the s-boson approach in the strongly mixed valent regime discussed above are a consequence of the unnormalized $s$-boson.  The normalization procedure of the $s$-boson may resolve this issue by projecting out unphysical states, potentially enabling accurate description of strong mixed valence regimes where Hubbard bands merge with the central Kondo resonance. 


All analytic slave particle mean-field formalisms of the SIAM exhibit a spurious mean-field transition temperature $T_c$ into the Kondo phase that does not exist in reality,  since the Kondo effect is a crossover. Our approach already demonstrates that the formation of the Kondo resonance is well-described in the ``normal state" where the s-boson is
uncondensed. This opens the possibilty that we can describe the development of coherence in heavy fermion lattice purely by using the Gaussian fluctuations of the s-field. 
It will also be interesting in future work to examine to what extent the temperature dependent normalization of the $s$-boson (Eq. \eqref{normy}) cures the \lhl{overestimated strongly mixed valent behavior}, and whether its feedback effect will \lhl{ameliorate} the issue of the finite temperature phase transition.

Lastly, it is apparent that this method can be generalized  to more complex atomic limits involving spin-orbit coupling and multiorbital physics with intra-orbital, inter-orbital, and Hund's interactions. Further directions encompass systems with more complicated conduction bath density of states and hybridization functions, as well as applications to the periodic Anderson and Hubbard models. 

\begin{acknowledgements}
The authors are grateful for useful conversations with A. Gleis and K. Roy. This work was supported by the Office of Basic Energy Sciences, Material Sciences and Engineering Division, U.S. Department of Energy (DOE) under contract DE-FG02- 99ER45790 (L.L.H.L. and P.C.).
\end{acknowledgements}
\bibliography{combined7,twodimensionheterostructures, andersonholstein}
\appendix
\begin{widetext}
\section{Calculation of the Effective Action}\label{sec:appendixeffectiveaction}
We start with the action for the SIAM with the physical $d$-electron represented as  $d\dg_\sigma \rightarrow f\dg_\sigma \left(b + t\dg\right)$, where $f\dg_\sigma$ is a spinon operator and the two bosons $b$ and $t$ keep track of valence fluctuations,
\begin{eqnarray}\label{eq:appSA0}
   S_A = \int_0^{\beta} d \tau \left[f\dg_{\sigma} (\partial_{\tau} + \lambda) f_{\sigma} + \bar{b} (\partial_{\tau} + \Delta E^Q_- + \lambda) b+ \bar{t} (\partial_{\tau} + \Delta E^Q_+ - \lambda) t + \frac{V_0}{\sqrt{N}} \sum_{\bk \sigma} \left(c\dg_{\bk \sigma} f_{\sigma} \left(\bar{b} + t \right) + \text{h.c.} \right) - \lambda Q\right],
\end{eqnarray}
where the atomic ionization energies are,
\begin{eqnarray}\label{eq:appionizationenergies}
    \Delta E^Q_{+}  &=& E_{Q+1} - E_Q = \tilde{\epsilon}_d + U/2 \cr 
    \Delta E^Q_{-} &=& E_{Q-1} - E_Q = -\tilde{\epsilon}_d + U/2.
\end{eqnarray}
where we have introduced the notation
\begin{eqnarray}
    \tilde{\epsilon}_d = \epsilon_d + U (Q-N/2)
\end{eqnarray} for the $d$-level position at doping away from half filling. 
We can write a symmetric $s$ and antisymmetric $a$ boson:
\begin{equation}\label{eq:bosons}
   s = (b + \bar{t}), \quad a = (b - \bar{t}),
\end{equation}
such that the particle hole transformation acts as $s \rightarrow \bar{s}$, and $a \rightarrow - \bar{a}$.
\begin{eqnarray}
    S_A &=& \int_0^{\beta} d \tau \left[f\dg_{\sigma} \left(\partial_\tau  +\lambda \right)f_{\sigma} + \frac{1}{4}(\bar{s} + \bar{a}) (\partial_{\tau} -\tilde{\epsilon}_d + U/2 + \lambda )(s + a) \right.\cr  &+&  
    \frac{1}{4}(s - a )(\partial_{\tau} + \tilde{\epsilon}_d + U/2  -\lambda) (\bar{s} - \bar{a}) + \left.\frac{V_0}{\sqrt{N}}\sum_{\bk \sigma} \left(c\dg_{\bk \sigma} f_\sigma \bar{s} + \text{h.c.}\right) - \lambda Q\right],
\end{eqnarray}
We can commute $\bar{t} t = t \bar{t}$, which are just numbers, which flips of the sign for $\partial_\tau$, which can be seen more clearly in Matsubara frequency.
\begin{eqnarray}
    S_A &=& \int_0^{\beta} d \tau \left[f\dg_{\sigma} \left(\partial_\tau  + \lambda\right) f_{\sigma} + \frac{1}{4}(\bar{s} + \bar{a}) (\partial_{\tau}  + U/2 + \lambda-\tilde{\epsilon}_d )(s + a) \right.\cr  &+&  \frac{1}{4}(\bar{s} - \bar{a} )(-\partial_{\tau}  + U/2  -\lambda + \tilde{\epsilon}_d) (s - a) + \left.\frac{V_0}{\sqrt{N}}\sum_{\bk \sigma} \left(c\dg_{\bk \sigma} f_\sigma \bar{s} + \text{h.c.}\right) - \lambda  Q\right],
\end{eqnarray}
Expanding,
\begin{eqnarray}\label{eq:expanded}
    S_A &=& \int_0^{\beta} d \tau \left[f\dg_{\sigma} \left(\partial_\tau  + \lambda  \right) f_{\sigma} +  \frac{U}{4}(\bar{s}s + \bar{a}a) \right.  +  \frac{1}{2}\bar{s}\left(\partial_\tau  + \lambda - \tilde{\epsilon}_d \right) a \cr &+& \frac{1}{2}\bar{a}\left(\partial_\tau  + \lambda -\tilde{\epsilon}_d\right)s  + \left.\frac{V_0}{\sqrt{N}}\sum_{\bk \sigma} \left(c\dg_{\bk \sigma} f_\sigma \bar{s} + \text{h.c.}\right) - \lambda Q\right].
\end{eqnarray}
Only the $s$ boson couples to the fermions, so we can integrate out the $a$ bosons to get,
\begin{eqnarray}\label{eq:appSA2}
   S_A  &=& \int_0^{\beta} d \tau \left[\bar{f}_{\sigma} \biggl(\partial_{\tau} + \lambda \biggr) f_{\sigma} +  \frac{1}{U}\bar{s}
   \left(\left(\frac{U}{2}\right)^2 - {(\partial_{\tau}  + \lambda - \tilde{\epsilon}_d )^2} \right) s + \frac{V_0}{\sqrt{N}} \sum_{\bk \sigma} \left(c\dg_{\bk \sigma} f_{\sigma} \bar{s} + \text{H.c.} \right)- \lambda Q \right],
\end{eqnarray}
\lhl{
\section{Atomic Limit Results}\label{sec:atomiclimitapp}
Let us first write down the effective action in the atomic limit,
\begin{equation}\label{eq:atomicSapp}
  S_{\text{atom}}= \int_0^{\beta}d\tau \left[
     \frac{1}{U}\bar{s}
   \biggl(\frac{U^2}{4\ }- (\partial_{\tau}  + \lambda -\tilde{\epsilon}_d )^2 \biggr) s +\sum_\sigma f\dg_{\sigma} (\partial_{\tau} + \lambda) f_{\sigma} - \lambda Q\right].
\end{equation}
The physical $d$-electron is the product of the $d$-spinon and the $s$-boson,
\begin{equation}\label{eq:physicaldelectronapp}
    d\dg_\sigma(\tau)= f\dg_{\sigma}(\tau) s(\tau).
\end{equation}
We can then factorize the physical $d$-electron Green's function in the atomic limit,
\begin{eqnarray}\label{eq:delectrongreensapp}
    G^{at}_d(\tau) &=& -\langle \mathcal{T} d_\sigma (\tau) d\dg_\sigma(0)\rangle = -\langle \mathcal{T} s\dg(\tau) f_\sigma (\tau) f\dg_\sigma(0) s(0) \rangle \cr &\approx&-\langle \mathcal{T}  f_\sigma (\tau) f\dg_\sigma(0) \rangle \langle s\dg(\tau) s(0) \rangle \cr &=& - G_{f}(\tau) D^{at}_s(-\tau),
\end{eqnarray}
where the $s$-boson Green's function in the atomic limit is only given by its bare kinetics. The bare $s$-propagator is,
\begin{equation}
    D_s^{at}(i\nu_m)^{-1}  = - \frac{1}{U} \left[\left(\frac{U}{2}\right)^2 - \left( \lambda - \tilde \epsilon_d - i\nu_m \right)^2 \right],
\end{equation}
with two poles at $\omega_+ = \Delta E^{Q}_+ + \lambda =  U/2  - \tilde \epsilon_d + \lambda$ and $\omega_-  = -\Delta E^{Q}_- -\lambda = -U/2  + \tilde \epsilon_d - \lambda$.
Using the convolution theorem, 
\begin{equation}\label{eq:bosonicmatsubarasumapp}
    G_d^{at}(i\omega_n) = -\frac{1}{\beta} \sum_{i\nu_r} G_{f}(i\omega_n + i\nu_r)D_s^{at}(i\nu_r),
\end{equation}
where the $f$-spinon propagator is,
\begin{equation}
    G_f (i \omega_n) = \left(i\omega_n - \lambda\right)^{-1}.
\end{equation}
Performing the bosonic Matsubara sum \label{eq:bosonicmatsubarasumapp} as an integral on the complex plane,
\begin{eqnarray}
    G_d(i\omega_n) &=&   \ointclockwise_C \frac{dz \; n_B(z)}{2 \pi i} G_f(i \omega_n + z) D_s(z) \cr &=& \int \frac{dv}{\pi}D^{''}_s(v - i\delta) \left[\frac{n_B(v) +q}{i \omega_n -\lambda + v}\right],
\end{eqnarray}
where the contour $C$ is taken to be clockwise around the poles and branch cuts of $G_f$ and $D^{at}_s$. Here $n_B(\omega) = 1/ \left(e^{\beta \omega} - 1\right)$ is the Bose-Einstein distribution and we have used $n_B(\lambda - i \omega_n) = - f(\lambda) = \lhl{-}q$, where $f(\lambda)= 1/ (e^{\beta \lambda} +1) $ is the Fermi function and \lhl{$q = Q/N$ is the ratio of the integer number of $d$-electrons and the total number of $d$-electron flavors}.
The imaginary part of the $d$-electron propagator is then related to the spectral function,
\begin{eqnarray}\label{eq:spectralfunc}
    A_d^{at}(\omega) &=& G^{''}_d(\omega - i\delta)/\pi \cr &=& \frac{1}{ \pi}\left[n_B(\lambda  - \omega ) + q \right]\,  D^{at \, ''}_s(\lambda - \omega) \cr &\xrightarrow{T = 0}& \left(1 - \frac{Q}{N} \right)\, \delta\left(\omega - \Delta E^{Q}_+ \right) + \frac{Q}{N} \,\delta\left(\omega + \Delta E^Q_-\right),
\end{eqnarray}
}

\section{Mean-Field Theory Analytical Results}\label{sec:meanfieldanalytics}
We now present some analytical results for a single impurity Anderson model with a conduction sea with a constant density of states $\rho_c$. We will first derive the mean-field results using both the ``top'' $t$ and ``bottom'' $b$ bosons, before switching over to the more economical symmetric $s$ boson.

Starting from \lhl{the} saddle-point of the action \eqref{eq:appSA0}, where $b(\tau) = b$ and $t(\tau) = t$ have no dynamics, integrating out the conduction band yields the following free energy,
\begin{equation}\label{eq:FMFapp1}
    F = -T N \sum_{|\omega_n| < D_c} \ln{\left[\lambda - i \omega_n + \frac{1}{N} \sum_{\bk}\frac{V_0^2(\bar b + t)(b + \bar{t})}{i\omega_n - \epsilon_\bk} \right]} + (\Delta E^Q_+ - \lambda) \; \bar{t} t + ( \Delta E^Q_- + \lambda) \; \bar{b} b - \lambda Q,
\end{equation}
where $D_c= \text{min}\left(D, U/2\right)$, the high-energy cutoff, depends on the relative size of the bandwidth $D$ and the Coulomb interaction $U$. If we consider a conduction sea with a constant density of conduction states $\rho_c$ with a band-width $2D$, the mean-field free energy can be rewritten as,
\begin{equation}\label{eq:FMFapp2}
    F = -T N \sum_{|\omega_n| < D_c} \ln{\left[\lambda - i \omega_n -i\tilde \Delta_n \right]} e^{i \omega_n 0^+} + (\Delta E^Q_+ - \lambda) \; \bar{t} t + ( \Delta E^Q_-+\lambda) \; \bar{b} b - \lambda Q,
\end{equation}
where we have explicitly written out the convergence factor and the $f$-electron self energy due to the hybridization with the conduction sea can be approximated as,
\begin{equation}
    \frac{1}{N} \sum_{\bk}\frac{V_0^2(\bar b + t)(b + \bar{t})}{i\omega_n - \epsilon_\bk} \approx -i\tilde\Delta \; \rm{sgn}(\omega_n),
\end{equation}
where the real part is negligible due to being $\mathcal{O}(\omega_n/D)$ and taking the large bandwidth approximation. Here we define, $\tilde{\Delta} = \pi \rho_c V_0^2(\bar b + t)(b + \bar{t}) /N  = \Delta (\bar b + t) (b+ \bar t)/N$, where $\Delta$ is the large-$N$ bare hybridization width, and $\tilde \Delta_n \equiv \tilde \Delta \;\rm{sgn}(\omega_n)$. 

Carrying out the particle-hole transformation on the free energy,
\begin{eqnarray}\label{eq:FMFapp2particlehole}
    F &=& -T N \sum_{|\omega_n| < D_c} \ln{\left[-\lambda - i \omega_n -i\tilde \Delta_n \right]} e^{i \omega_n 0^+} + (\Delta E^Q_- + \lambda) \; \bar{b} b + ( \Delta E^Q_+-\lambda) \; \bar{t} t - \lambda (Q - N) \cr &=& -T N \sum_{|\omega_m| < D_c} \ln{\left[\lambda - i \omega_m -i\tilde \Delta_m \right]} e^{-i \omega_m 0^+} + (\Delta E^Q_- + \lambda) \; \bar{b} b + ( \Delta E^Q_+-\lambda) \; \bar{t} t - \lambda (Q - N),
\end{eqnarray}
where we have relabeled $\omega_m = - \omega_n$. Summing \eqref{eq:FMFapp2particlehole} and \eqref{eq:FMFapp2}, we obtain,
\begin{equation}\label{eq:FMFapp2particlehole2}
     F = -T N \sum_{|\omega_n| < D_c} \ln{\left[\lambda - i \omega_n -i\tilde \Delta_n \right]} \cos{\left(\omega_n 0^+\right)} + (\Delta E^Q_+ - \lambda) \; \bar{t} t + ( \Delta E^Q_-+\lambda) \; \bar{b} b - \lambda \left( Q - N/2\right),
\end{equation}

Introducing the complex $d$-level $\xi = \lambda + i\tilde \Delta$, then
\begin{eqnarray}
    \sum_{|\omega_n| < D_c} \ln{\left[\lambda - i \tilde \Delta_n - i\omega_n\right]} \cos{\left(\omega_n0^+\right)} &=& 2 \text{Re} \sum_{D_c > \omega_n > 0 } \ln{\left[\xi + i \omega_n \right]} \cr &=& 2 \text{Re} \sum_{n = 0}^{M-1} \ln{\left[n + \frac{1}{2} + \frac{\xi}{2 \pi iT } \right]} + \rm{const} \cr &=& 2 \text{Re}  \ln{\left[\prod_{n = 0}^{M-1}\left(n + \frac{1}{2} + \frac{\xi}{2 \pi iT }\right) \right]} + \rm{const},
\end{eqnarray}
where we have introduced $M = D_c /(2 \pi T)$, and we will ignore the overall constant, proportional to $\ln{\left[2 \pi i T\right]}$ on the second line. Since the gamma function $\Gamma(x)$ has the property $\Gamma(1+x) = x \Gamma (x)$, it follows that,
\begin{equation}
    \prod_{n = 0}^{M-1}  (n+x)\Gamma(x) = \Gamma(M+x),
\end{equation}
so that
\begin{equation}
    \prod_{n = 0}^{M-1} (n+x) = \frac{\Gamma(M+x)}{\Gamma(x)}.
\end{equation}
Hence, we can rewrite, 
\begin{equation}
    \prod_{n = 0}^{M-1}\left(n + \frac{1}{2} + \frac{\xi}{2 \pi iT }\right) = \frac{\tilde \Gamma\left(\xi + i D_c\right)}{\tilde \Gamma(\xi)},
\end{equation}
where we have introduced the short-hand,
\begin{equation}
    \tilde \Gamma (z) = \Gamma \left(\frac{1}{2} + \frac{z}{2 \pi i T} \right).
\end{equation}
We can now rewrite the mean-field free energy \eqref{eq:FMFapp2} in the compact form,
\begin{eqnarray}\label{eq:FMFapp3}
    F &=& -2 T N \text{Re} \ln{\left[\frac{\tilde \Gamma(\xi + i D_c)}{\tilde \Gamma(\xi)}\right]}+ (\Delta E^Q_+ - \lambda) \; \bar{t} t + ( \Delta E^Q_-+ \lambda) \; \bar{b} b - \lambda \left(Q - N/2\right) \cr &=& N \text{Re}\left(2 T \ln \tilde \Gamma(\xi) - \frac{\xi}{i \pi} \ln{\left[\frac{D_c}{2 \pi T}\right]}\right) + (\Delta E^Q_+ ) \; \bar{t} t + ( \Delta E^Q_-) \; \bar{b} b - \lambda \left( Q - N/2 \right),
\end{eqnarray}
where we have assumed $\xi$ is small compared to $D_c$ to \lhl{Taylor} expand,
\begin{eqnarray}
    2T\ln{\tilde \Gamma(\xi + i D_c)} &\approx& 2T \ln{\tilde \Gamma(i D_c)} + \tilde{\psi}(iD_c) \frac{\xi}{i \pi} \cr &\approx& 2T \ln{\tilde \Gamma(i D_c)} + \frac{\xi}{i \pi} \ln{\frac{D_c}{2 \pi T}}
\end{eqnarray}
where we have denoted $\tilde \psi(\xi) = \psi\left(\frac{1}{2} + \frac{\xi \beta}{2 \pi i}\right)$, and $\psi(z) = \frac{\partial \ln{\Gamma(z)}}{\partial z}$ is the digamma function. The second line, we have assumed $D_c/(2\pi T)$ is large and replaced the digamma function's asymptotic behavior with a logarithm, $\psi(\frac{1}{2} + z) \rightarrow \ln{z}$ for large $z$.

The saddle-point equations can be found by differentiating with respect to $\lambda$, $b$, and $t$. First differentiating with respect to $\lambda$,
\begin{eqnarray}\label{eq:FMFapp4}
\frac{\partial F}{\partial \lambda} &=& \frac{N}{\pi} \text{Im} \left[\tilde \psi(\xi)\right] + \bar{b}b - \bar t t-Q +N/2= 0 \cr &=&  \frac{N}{\pi} \text{Im} \left[\tilde \psi(\xi) + \frac{i \pi}{2} \right] + \bar{b}b - \bar t t-Q = 0. 
\end{eqnarray}
Differentiating with respect to the two bosons yields, 
\begin{eqnarray}\label{eq:FMFapp5}
    \frac{\partial F}{\partial \bar t} &=& \frac{N\Delta (\bar b + t)}{\pi} \text{Re}\left[\tilde \psi(\xi) - \ln{\left(\frac{D_c}{2 \pi T}\right) }\right] + \left(\Delta E^Q_+ - \lambda\right) t = 0, \cr \frac{\partial F}{\partial \bar b} &=& \frac{N\Delta ( b + \bar t)}{\pi} \text{Re}\left[\tilde \psi(\xi) - \ln{\left(\frac{D_c}{2 \pi T}\right) }\right] + \left(\Delta E^Q_- +\lambda\right) b = 0.
\end{eqnarray}
We can hence see that we could have absorbed $\lambda N/2$ into an additional $i$ in the logarithm and rewritten the free energy as,
\begin{equation}\label{eq:Fapprewritten}
   F =  N \text{Re}\left(2 T \ln \tilde \Gamma(\xi) - \frac{\xi}{i \pi} \ln{\left[\frac{D_c}{2 \pi i T}\right]}\right) + (\Delta E^Q_+ ) \; \bar{t} t + ( \Delta E^Q_-) \; \bar{b} b - \lambda Q.
\end{equation}

Alternatively, the mean-field free energy can be derived using the saddle-point approximation for the action \eqref{eq:appSA2} where the $s$-boson has no dynamics. Integrating out the conduction band yields,
\begin{equation}\label{eq:analyticmfFapp}
    F = -TN \sum_{\omega_n < D_c} \ln{\left[\lambda - i \tilde \Delta_n - i\omega_n\right]} + \frac{\vert s \vert^2 V_0^2}{J(\lambda)} - \lambda Q,
\end{equation}
where $i \tilde\Delta_n = i \pi \rho_c V_0^2 \vert \tilde s\vert^2 \text{sgn}(n)$ is the large-cutoff approximation for the f-electron self energy, where $\tilde s = s/\sqrt{N}$. Lastly, the  $\lambda$-dependent Kondo coupling has the Schrieffer-Wolff form,
\begin{equation}\label{eq:Jappendix}
    J(\lambda) = \frac{V_0^2}{U/2 + \lambda - \tilde \epsilon _d }+ \frac{V_0^2}{U/2 - \lambda + \tilde \epsilon _d }
\end{equation}
We can rewrite the free energy \eqref{eq:analyticmfFapp},
\begin{eqnarray}\label{eq:analyticmfFapp2}
    F &=& N \text{Re}\left(2 T \ln \tilde \Gamma(\xi) - \frac{\xi}{i \pi} \ln{\left[\frac{D_c}{2 \pi i T}\right]}\right) +  \frac{\vert s \vert ^2 V_0^2}{J(\lambda)} - \lambda Q \cr &=& N \text{Im} \left[2 i T \ln{\tilde \Gamma(\xi)}- \frac{\xi}{\pi} \ln{\frac{D_c}{2 \pi i T}} \right] +  \frac{\vert s \vert ^2 V_0^2}{J(\lambda)} - \lambda Q.
\end{eqnarray}
We can then rewrite,
\begin{eqnarray}
    \frac{\vert s \vert ^2 V_0^2}{J(\lambda)} - \lambda Q &=& \frac{N\; \text{Im}\left[\xi\right]}{\pi}\left[ \ln{e^\frac{1}{\rho_c J(\lambda)} } - \ln e^{i \pi q} \right],
\end{eqnarray}
where $q = Q/N$. Hence the free energy is,
\begin{eqnarray}\label{eq:analyticmfFapp3}
    F &=&    N \;\text{Im} \left[2 i T \ln{\tilde \Gamma(\xi)}- \frac{\xi}{\pi} \ln{\frac{T_K(\lambda) e^{i \pi q}}{2 \pi i T}} \right] \equiv  N \;\text{Im} F_c \;,
\end{eqnarray}
where $T_K(\lambda) = D_c \exp{\left[-\frac{1}{ \rho_c J(\lambda)}\right]}$ is the $\lambda$-dependent Kondo temperature.

In the Kondo limit, we can neglect the $\lambda$ dependence of the Kondo coupling constant, setting $J = J(\lambda = 0)$. \lhl{In this limit}, we obtain the saddle point equations by differentiating \eqref{eq:analyticmfFapp3} with respect to $\xi$. Setting $\partial F_c/\delta \xi = 0$,   we obtain 
\begin{equation}
    \tilde \psi(\xi) = \ln \left(\frac{T_K e^{i \pi q}}{2 \pi  i T}\right),
\end{equation}
$T_K = T_K(\lambda)\vert_{\lambda =0} =D_c \exp{\left[-\frac{1}{ \rho_c J}\right]} $ is the Kondo temperature.  At low temperature, we can approximate $\tilde \psi (z) = \ln (\frac{z}{2 \pi i T})$, so we obtain simply $\xi = \lambda + i \tilde \Delta = T_K e^{i \pi q}$, recovering the large $N$ limit of the SU(N) Kondo model, but with the finite $U$  Schrieffer-Wolff form of the coupling constant $J$\eqref{eq:Jappendix}.  

For the fully mixed valence regime, the $\lambda$ dependence in the Kondo coupling $J(\lambda)$ cannot be ignored, and the saddle-point equations are obtained by minimizing \eqref{eq:analyticmfFapp2}, replacing Eqs. \eqref{eq:FMFapp4} and \eqref{eq:FMFapp5},
\begin{eqnarray}
    \frac{\partial F}{\partial \lambda} &=& \frac{N}{\pi}\left(\text{Im}\left[\tilde \psi(\xi) - \ln{\frac{D_c}{2 \pi i T} }\right] +2 \pi \frac{(\tilde \epsilon_d - \lambda)}{U}\vert \tilde s\vert^2 -\pi q\right) = 0 \cr  &=& \frac{N}{\pi}\left(\text{Im}\left[\tilde \psi(\xi) + \frac{i \pi}{2} \right] +2 \pi \frac{(\tilde \epsilon_d - \lambda)}{U}\vert \tilde s\vert^2 -\pi q\right) = 0\cr
    \frac{\partial F}{\partial \bar s} &=& N s V_0^2 \rho_c \left(\text{Re} \left[ \tilde \psi(\xi) - \ln{\frac{T_K(\lambda)}{2 \pi T}} \right]\right) = 0,
\end{eqnarray}
The two saddle-point equations can be written as a single complex equation,
\begin{equation}\label{compappendix}
    \tilde\psi(\xi) - \ln{\left(\frac{T_K(\lambda) e^{i \pi q}}{2 \pi i T}\right)} = - 2 \pi i \frac{\tilde \epsilon_d-\lambda }{U} \vert \tilde s \vert^2.
\end{equation}

\section{Convexity and Solutions for $T_c$}\label{sec:convexity}
The mean-field equation for $T_c$,
\begin{equation}\label{eq:Tcsolveappendix}
    \text{Re}\left[\psi \left(\frac{1}{2} + \frac{\log{\left[1/q - 1\right]}}{2 \pi i}\right) - \ln\left(\frac{  T_K(T_c \log{\left[1/q - 1\right]})}{2\pi T_c}\right) \right] =0,
\end{equation}
can be rewritten as,
\begin{equation}\label{eq:generalf}
    f(T_c) = a T_c^2 + bT_c +c + \lhlnew{\ln{\left(\frac{D_c}{2 \pi T_c} \right)}} = 0,
\end{equation}
where
\begin{equation}
    a = \frac{\pi}{U \Delta} \left( \ln{\left[1/q - 1\right]} \right)^2, \quad b = -\frac{2 \pi \tilde \epsilon_d}{U \Delta}\ln{\left[1/q - 1\right]}, \quad c= \frac{\pi}{U \Delta}\left(\tilde \epsilon_d^2 - \left(\frac{U}{2}\right)^2\right)  - \textrm{Re} \left[\psi \left(\frac{1}{2} + \frac{\log{\left[1/q - 1\right]}}{2 \pi i}\right)\right].
\end{equation}
For the half-filled impurity $q = 1/2$, $a = b = 0$ and therefore there is only one solution,
\begin{equation}\label{eq:Tchalffillinganalyticapp}
    T_c^{q=1/2} = e^c = \frac{T_K(0)}{2 \pi} \exp{\left\{ - \psi\left(\frac{1}{2}\right)\right\}},
\end{equation}
where $\psi(1/2)$ is purely real and,
\begin{equation}
    T_K(0) = D_c \exp{\left[-\frac{\pi}{U \Delta} \left(\left(\frac{U}{2}\right)^2 - \tilde \epsilon_d^2\right)\right]}.
\end{equation}
For $a > 0$, $f(T_c)$ is convex,
\begin{equation}\label{eq:convexshow}
    f^{''}(T_c) = 2a + \frac{1}{T_c^2} \geq 0, \quad T_c \geq 0,
\end{equation}
where equality in \eqref{eq:convexshow} only occurs in the unphysical $T_c \rightarrow \infty$ limit. Hence $f(T_c)$ is convex and has either two solutions, one solution, or no solutions. The minimum at $T_c = T_*$ is when $f(T_c)$ has a turning point,
\begin{equation}\label{eq:minimumshow}
    f^{'}(T_c)\rvert_{T_c = T_*} = 2 a T_* + b - \frac{1}{T_*} = 0.
\end{equation}
Solving this equation for the minimum for a positive temperature,
\begin{equation}\label{eq:Tstar}
    T_* = \frac{-b + \sqrt{b^2 + 8a}}{4a}.
\end{equation}
The mean-field equation \eqref{eq:Tcsolveappendix} has two solutions if $f(T_*) < 0$, one solution if $f(T_*) = 0$, and no solutions if $f(T_*) > 0$.

We want to know for what values of the $d$-level $\tilde \epsilon_d$, which affect the variables $b$ and $c$ in \eqref{eq:generalf}, there are no solutions for $T_c$. We can do so by finding the critical $d$-level $\tilde \epsilon_{d,*}$ where we only get one solution, which is when $f(T_*) = 0$, which can be done via a root finding algorithm.

As we did in the main text, but now plotting all the $T_c$ solutions to \eqref{eq:Tcsolveappendix} rather than just the lowest $T_c$ values, we plot the transition temperatures $T_c$ into the mean-field Kondo phase (Fig. \ref{fig:Tcappendix}) as a function of electron filling $x$ for an $N = 6$ SIAM lith a bare hybridization width in the large-$N$ analysis of $\Delta = \pi \rho_c V_0^2 = 25$, onsite Hubbard $U = \lhlnew{4\Delta}$, and a flat conduction density of states with \lhlnew{bandwidth $2D$ and a cutoff $D = 4\Delta/5$}.

We mimic the resetting of the Coulomb blockade physics (when reaching the next impurity valence) by manually controlling two parameters as a function of $x$. The electron filling $x$ is taken to be a continuous variable where $x = 0$ when the impurity is empty, and $x = 6$ is when the impurity is completely full. First, we take $Q$, the number of $d$-electrons in the atomic limit of the impurity, to increase stepwise as a function of $x$,
\begin{equation}\label{eq:mftqcontrolapp}
Q(x) =     \lfloor x \rceil \in \mathbf{Z}^+,
\end{equation}
and only taking positive integer values by taking the rounded value of $x$. For example, in the range $w\leq x < 2.5$, $Q = 2$, and $Q = 3$ when $2.5 \leq x <3.5$. For each integer $Q$, we produce a separate curve (either $T_c$ or $\langle n_f \rangle $) where we linearly vary the $d$-level $\tilde \epsilon_d$ and reset the value of $\tilde \epsilon_d$ for the next integer $Q$,
\begin{equation}\label{eq:mftEdcontrolapp}
   \tilde \epsilon_d(x) =  -U \left(x - Q(x) \right).
\end{equation}
$\tilde \epsilon_d$ equals $U/2$ at the half-integer below integer $x$, linearly decreasing to $U/2$ at the half-integer above integer $x$, with $\tilde \epsilon_d = 0 $ at integer $x$. This resets for the next integer $Q$ value, running over the next $x$ range. For a given integer $Q(x)$, certain $\tilde \epsilon_d(x)$ values may not have a solution for $T_c$ so $x$ will be restricted for each integer $Q$ in a way to not include the values of $\tilde \epsilon_d(x)$ which have no transition temperature solution, as discussed above.

\begin{figure}[t]
       \centering
       \includegraphics[width=0.5\linewidth]{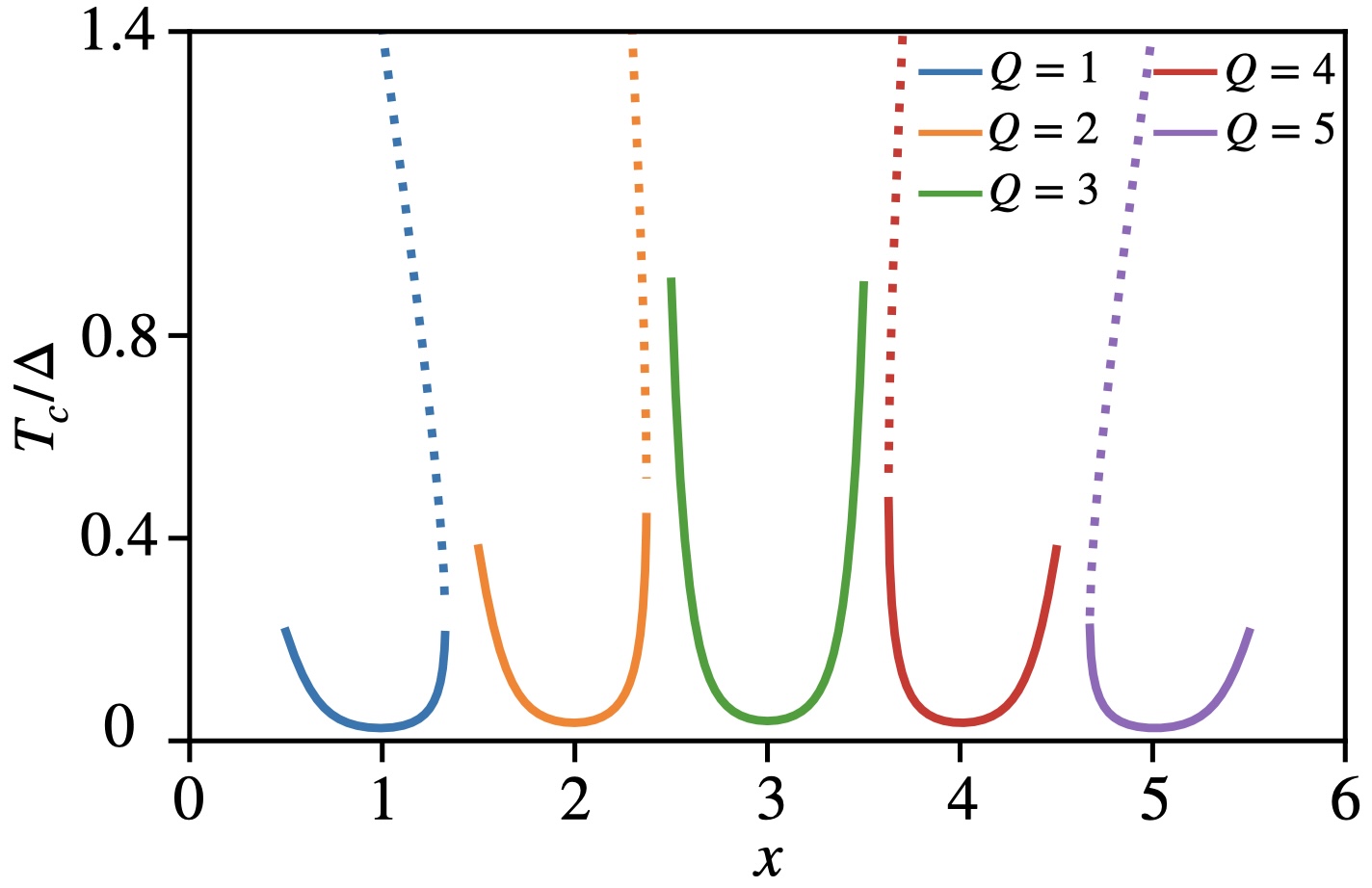}
       \caption{Mean-field transition temperature $T_c$ \lhlnew{in units of $\Delta$} plotted versus electron filling $x$ for $\Delta = 25$, $U = \lhlnew{4\Delta}$, and a conduction bath \lhlnew{with bandwidth $2D$ and cutoff $D = 4\Delta/5$} for different integer atomic fillings $Q$ of the $N = 6$ impurity ($Q = 1$, blue, $Q = 2$, orange, $Q = 3$, green, $Q = 4$, red, and $ Q = 5$, purple). For each $Q$, the $d$-level position from half filling $\tilde \epsilon_d$ is taken to vary linearly from $U/2$ to $-U/2$ between half-integer fillings. $x$ is restricted for each $Q$ to the range where there is at least one solution for $T_c$. The solid lines are the lower calculated $T_c$ and the dotted lines are the higher temperature $T_c$ solution for the re-entrant Kondo mean-field phase.}  
       \label{fig:Tcappendix}
\end{figure}

\section{Gaussian Fluctuations in the Cartesian Gauge}\label{sec:gausscartesian}
The full expressions for the normal and anomalous Bose self energies are,
\begin{eqnarray}\label{chi0suscep}
    \Pi_0(i\nu_m) &=&  \left(\frac{\Delta}{\pi}\left[\tilde{\psi}(\xi + i \nu_m) - \tilde{\psi}(\xi)\right] \right)^* + \left(\frac{2 \tilde{\Delta}^2}{\nu_n (\nu_n + 2 \tilde{\Delta})}\right) \frac{\Delta}{\pi} \,\textrm{Re}\left[\tilde{\psi}(\xi + i \nu_m) - \tilde{\psi}(\xi)\right] + \frac{\Delta}{\pi}\, \textrm{Re} \left[\tilde{\psi}(\xi) - \ln{\frac{D}{2 \pi i T}} \right] \cr 
    \Pi_A(i \nu_m) &=& \frac{2 \tilde{\Delta}\left(\tilde{\Delta} + \nu_m\right)}{\nu_m(\nu_m + 2 \tilde{\Delta})}  \frac{\Delta}{\pi} \, \textrm{Re} \left[\tilde{\psi}(\xi + i \nu_m) - \tilde{\psi}(\xi)\right],
\end{eqnarray}
where the large-$N$ bare resonance width is $\Delta = \pi \rho_c V_0^2$ and $\rho_c$ is the constant conduction bath density of states at the Fermi level. $\tilde{\Delta} = \Delta \vert\tilde{s}\vert ^2$ is the renormalized resonance width when the $s$-boson gains a mean-field expectation value $\tilde{s}$. The Bose self-energies above are related to the corresponding susceptibilities by a negative sign. The complex $d$-level position is given by $\xi = \lambda  + i\tilde{\Delta} = \lambda $, where the imaginary part vanishes because the renormalized hybridization width $\tilde{\Delta} = \Delta \vert s\vert^2$ equals zero in the normal state when the $s$-boson has yet to develop a mean-field expectation value. In the normal boson state, the anomalous self energy $\Pi_A(i\nu_m)$ and the central term in the normal self energy $\Pi_0$ also both vanish.

\end{widetext}

\end{document}